\newif\ifconfver
\confvertrue        %
\RequirePackage{fix-cm}
\ifconfver
\documentclass[10pt,journal,final,twoside]{IEEEtran}
\else
\documentclass[11pt,draftcls,onecolumn]{IEEEtran}
\fi

\usepackage{cite}
\usepackage[T1]{fontenc}
\usepackage[utf8]{inputenc}
\usepackage{times}
\usepackage{microtype}
\usepackage{amsmath,amssymb,amsfonts,mathtools,bm}
\usepackage{empheq}
\usepackage{booktabs}
\usepackage{multirow}
\usepackage{array}
\usepackage{makecell}   %
\usepackage{tabularx}
\usepackage{enumitem}
\usepackage{graphicx}
\usepackage[caption=false,font=footnotesize]{subfig} %
\usepackage{xcolor}
\usepackage{url}
\usepackage{algorithm}
\usepackage{algpseudocode}
\usepackage[hidelinks]{hyperref}
\usepackage{tikz}
\usetikzlibrary{arrows.meta,positioning,calc,backgrounds,decorations.pathreplacing,fit,patterns}
\newsavebox{\figcolbox}          %
\newsavebox{\figleftbox}         %
\newlength{\figcolht}            %
\usepackage{pgfplots}            %
\pgfplotsset{compat=1.18}
\usepgfplotslibrary{groupplots}   %
\usepgfplotslibrary{fillbetween}  %
\usepackage{pgfplotstable}         %
\definecolor{pcol}{HTML}{0072B2}
\definecolor{dcol}{HTML}{E69F00}
\definecolor{gcol}{HTML}{009E73}
\definecolor{vcol}{HTML}{D55E00}
\definecolor{ink}{HTML}{243240}
\definecolor{slate}{HTML}{6E7C8A}
\tikzset{
  io/.style={draw,rounded corners=5pt,align=center,inner sep=3pt,font=\footnotesize,fill=black!4},
  blk/.style={draw,rounded corners=2pt,align=center,inner sep=3pt,font=\footnotesize,fill=blue!5},
  op/.style={draw,rounded corners=3pt,align=left,inner sep=5pt,font=\footnotesize,fill=blue!4},
  lossb/.style={draw,rounded corners=3pt,align=center,inner sep=4pt,font=\footnotesize,fill=orange!10},
  flow/.style={-{Latex[length=2mm]},semithick},
  back/.style={-{Latex[length=2mm]},semithick,dashed},
  note/.style={font=\scriptsize,align=center}
}

\allowdisplaybreaks[3]
\input{mysymbol.sty}
\newtheorem{remark}{\bfseries Remark}
\makeatletter
\def\@begintheorem#1#2{\@IEEEtmpitemindent\itemindent\relax\topsep 0pt\rmfamily\trivlist%
    \item[]\textit{\noindent #1\ #2:} \itemindent\@IEEEtmpitemindent\relax}
\def\@opargbegintheorem#1#2#3{\@IEEEtmpitemindent\itemindent\relax\topsep 0pt\rmfamily \trivlist%
    \item[]\textit{\noindent #1\ #2\ (#3):} \itemindent\@IEEEtmpitemindent\relax}
\makeatother

\definecolor{mohamadcol}{HTML}{000000} %
\definecolor{mohamadhlcol}{HTML}{CDEFEC}
\newcommand{\mohamad}{\color{mohamadcol}{}}
\usepackage{soul}

\usepackage[normalem]{ulem}
\newcommand{\mohamadboxcite}[1]{\mbox{\mohamadorigcite{#1}}}
\DeclareRobustCommand{\mohamaddel}[1]{{\color{mohamadcol}\let\mohamadorigcite\cite\let\cite\mohamadboxcite\sout{#1}}}

\newcommand{\R}{\mathbb{R}}

\def\bbtau{{\bm{\tau}}}
\def\bbsigma{{\bm{\sigma}}}
\def\bbtheta{{\bm{\theta}}}
\newcommand{\prox}{\mathrm{prox}}

\newcommand{\diag}{\mathrm{diag}}
\newcommand{\Tmap}{\mathcal{T}}
\newcommand{\Jssl}{\mathcal{J}_{\mathrm{ssl}}}

\makeatletter
\def\subsection{\@startsection{subsection}{2}{\z@}{2ex plus 0.5ex minus 0.5ex}%
{0.7ex plus .5ex minus 0ex}{\normalfont\normalsize\itshape}}
\makeatother

\begin{document}

\title{\huge \textbf{Loop-OPF}: Solver-Informed Learning for Optimal Power Flow with Inference-Time Topology Adaptivity}

\author{
\IEEEauthorblockN{Mohamad Chehade},~\IEEEmembership{Student Member, IEEE}, and \IEEEauthorblockN{Hao Zhu},~\IEEEmembership{Senior Member, IEEE}

\thanks{\protect\rule{0pt}{3mm} This work has been supported by NSF Grant 2150571 and US Army Research Office (ARO). The authors acknowledge the Texas Advanced Computing Center (TACC) at The University of Texas at Austin for providing computational resources that have contributed to the research results reported within this paper.}
\thanks{\protect\rule{0pt}{3mm} The authors are with the Chandra Family Department of Electrical \& Computer Engineering, The University of Texas at Austin, Austin, TX 78712, USA; Emails: {\{chehade, haozhu\}{@}utexas.edu}.}
}

\markboth{PREPRINT}%
{Chehade \MakeLowercase{\textit{et al.}}: \textbf{Loop-OPF}: Solver-Informed Learning for Optimal Power Flow with Inference-Time Topology Adaptivity}

\maketitle

\begin{abstract}
Learning for optimal power flow (OPF) has recently been advocated to accelerate OPF solutions by using a pre-trained neural network (NN) to predict directly from given inputs such as the load demand. However, existing learning-for-OPF methods largely fall short of ensuring the optimality and feasibility of their predictions. They also struggle to adapt at inference time to unseen topology scenarios, such as those caused by line or generator outages. This work puts forth \textit{Loop-OPF}, a solver-informed NN that builds its layers from the iterations of the primal--dual hybrid gradient (PDHG) method. Under standard convexity and regularity conditions on DC-OPF, the resultant fixed-point output of Loop-OPF is thus optimal and feasible. To speed up PDHG across diverse OPF scenarios, Loop-OPF learns certain PDHG parameters in a self-supervised manner, without any OPF solution labels. We adopt the deep equilibrium (DEQ) framework for training so that the  memory cost does not grow with the number of layers. As the grid topology enters every layer as an input, the same trained Loop-OPF directly adapts to unseen contingencies without retraining. Numerical tests on the IEEE 300-bus and PEGASE 1354-, 2869-, and 9241-bus networks have demonstrated the accuracy and topology adaptivity of the proposed Loop-OPF over competing learning-for-OPF methods.
\end{abstract}

\begin{IEEEkeywords}
Optimal power flow, contingency analysis, solver-informed learning, deep equilibrium models, primal--dual hybrid gradient, self-supervised learning.
\end{IEEEkeywords}

\section{Introduction}
\label{sec:intro}

\IEEEPARstart{L}{earning-based} optimal power flow (OPF), or learning-for-OPF methods~\cite{kotary2021survey, fioretto2020acopf, singh2021sidnn}, have recently been advocated to accelerate the OPF solution for real-time grid operations. The main idea is to train a neural network (NN) model that can \textit{directly predict} the OPF solution from given problem inputs such as load demand. %
{\mohamad The growing body of learning-for-OPF literature has advanced the NN architecture~\cite{pan2019deepopf, falconer2023topology, liu2023topognn}, the loss function~\cite{fioretto2020acopf, singh2021sidnn, donti2021dc3, park2023pdl, chen2023e2elr, kim2024deeplde}, and the input--output mapping~\cite{pan2019deepopf, misra2022activesets, chatzos2021spatial}. Still, learning-for-OPF largely falls short in ensuring the optimality and feasibility of its predictions, and in adapting at inference to unseen topologies.} %
Addressing these existing gaps is imperative to accelerate the practical deployment of learning-for-OPF in grid operations. 

Model training in learning-for-OPF is either \textit{supervised} by input--output pairs of offline OPF solutions~\cite{pan2019deepopf, huang2022deepopfv, chatzos2021spatial, falconer2023topology} or, in an \textit{unsupervised} manner, guided by the OPF objective cost and constraints~\cite{donti2021dc3, park2023pdl, huang2024deepopfngt}. In this sense, the model output generally has no guarantee of its \textit{optimality} or \textit{feasibility}, i.e., the satisfaction of power balance or line flow limits. Determining binding line constraints is particularly critical for OPF~\cite{misra2022activesets}{\mohamad,~\cite{ardakani2013umbrella, pineda2020screening}}, because violating even %
{\mohamad just one} line limit %
could yield a substantially different dispatch. 
A common correction therefore adds penalty terms for constraint violations to the training loss~\cite{fioretto2020acopf, liu2023topognn, huang2024deepopfngt}, as in \textit{physics-informed neural networks}~\cite{nellikkath2022pinn}, but these heuristics  cannot strictly enforce feasibility{\mohamad~\cite{venzke2021verification}}. Other approaches have addressed power balance by construction, through completion~\cite{donti2021dc3} or repair layers~\cite{chen2023e2elr}. %
{\mohamad Line flow limits are handled far less effectively than power balance: most methods only penalize their violations, in primal or dual (Lagrangian) form~\cite{chen2023e2elr, park2023pdl, kim2024deeplde}, and some omit them altogether~\cite{donti2021dc3}.} %
While a projection onto the constraint set can fully recover a feasible solution~\cite{pan2019deepopf, liang2024homeomorphic}, this requires solving another optimization problem of similar complexity to OPF at inference. %
{\mohamad Meanwhile, a constraint calibration approach by tightening  inequality constraints during training to approach feasible predictions~\cite{zhao2023preventive}, but the tightening has to exclude part of the feasible region and thus sacrifices some optimality.} %
{\mohamad Hence, existing methods, at the best, attempt to attain feasibility  either with OPF-like computation complexity at inference or at the cost of optimality.} %

Besides optimality %
{\mohamad and feasibility}, another common limitation of existing methods lies in the lack of adaptivity to a new topology{\mohamad, e.g., after line switching~\cite{fisher2008ots},} or OPF instance. Largely tied to the specific network topology and set of line parameters seen by training, they cannot represent any new OPF formulation even with a small change in the limit of one binding line. Indeed, some have attempted to address this issue by using training data from multiple contingency scenarios, in particular graph neural networks (GNNs) that embed the grid-topology structure~\cite{falconer2023topology, liu2023topognn}. However, such approaches cannot guarantee the effectiveness beyond training scenarios to any unseen ones, especially with a different network congestion pattern.
In short, existing learning-for-OPF models still require complex mechanisms to ensure feasibility, and their applicability to unseen topologies remains an open question.
Table~\ref{tab:litreview} provides a summary of these approaches in comparison with our proposed Loop-OPF.

\begin{table}[t]
\centering
\caption{Learning-for-OPF Methods Compared with Proposed Loop-OPF}
\label{tab:litreview}
\footnotesize
\setlength{\tabcolsep}{2pt}
\begin{tabular*}{\columnwidth}{@{}l@{\extracolsep{\fill}}cccc@{}}
\toprule
 & Power & Flow & Needs & New \\
Method & balance & limits & labels & topology \\
\midrule
DeepOPF~\cite{pan2019deepopf} & Exact & Projection & Yes & No \\
DC3~\cite{donti2021dc3} & Exact & --- & No & No \\
E2ELR~\cite{chen2023e2elr} & Exact & Soft & No & No \\
DeepLDE~\cite{kim2024deeplde} & Exact & Soft & No & No \\
PDL~\cite{park2023pdl} & Soft & Soft & No & No \\
Homeomorphic~\cite{liang2024homeomorphic} & Exact & Exact$^{\dagger}$ & Optional & No \\
GNN~\cite{owerko2020gnnopf} & --- & --- & Yes & No \\
Topology GNN~\cite{liu2023topognn} & --- & Soft & Yes & Partial \\
\midrule
\textbf{Loop-OPF} & {\mohamad\textbf{Exact}} & {\mohamad\textbf{Exact}} & \textbf{No} & \textbf{Yes} \\
\bottomrule
\multicolumn{5}{@{}p{\columnwidth}@{}}{\scriptsize Exact: by construction; Projection: via an extra solve; Soft: penalized in training; ---: not handled. $^{\dagger}$Given a valid learned mapping.}
\end{tabular*}
\end{table}

In this work, we present Loop-OPF, a learned solver-informed NN that can ensure %
{\mohamad the optimality and feasibility of its prediction} %
and adapt to unseen topologies.
A solver-informed design~\cite{monga2021unrolling, chen2022l2o} builds the Loop-OPF layers from the iterations of primal--dual hybrid gradient (PDHG)~\cite{chambolle2011pdhg}, a powerful optimizer for large-scale constrained optimization{\mohamad~\cite{applegate2021pdlp}} %
As a result, %
{\mohamad for any convex OPF such as the DC-OPF, the output of Loop-OPF is guaranteed to be optimal and feasible at convergence~\cite{chambolle2011pdhg, boyd2004convex}.} %
{\mohamad To speed up PDHG convergence across diverse OPF instances~\cite{pan2019deepopf}, Loop-OPF learns certain PDHG parameters as a deep equilibrium (DEQ) model~\cite{bai2019deq}, with fixed training memory regardless of the number of layers.} %
Moreover, the network topology can be directly incorporated into Loop-OPF layers, allowing the same trained Loop-OPF to represent any unseen topology without retraining. %
{\mohamad Loop-OPF is not the first} %
solver-informed NN built from the PDHG iterations, with PDHG-Net~\cite{li2024pdhgnet} developed to accelerate the linear programming solution. However, our Loop-OPF offers two important differences. First, PDHG-Net serves only as an initializer for another first-order solver, whereas Loop-OPF itself returns the solution directly.  {\mohamad Second, PDHG-Net is trained by backpropagating through all of its unrolled layers, so its training memory grows with the number of layers. Loop-OPF is instead trained as a DEQ, differentiating only at its fixed point, so its training memory does not depend on the number of iterations.} %

In short, our contributions are summarized as follows:
\begin{enumerate}[leftmargin=*]
\item \textbf{Solver-informed architecture with guaranteed optimality.} Loop-OPF embeds the PDHG iterations in its architecture, so that its converged output is guaranteed to satisfy the optimality conditions of {\mohamad the} OPF problem (Section~\ref{sec:optimality}).
\item \textbf{Topology adaptivity.} Using diverse topology scenarios and OPF instances, the single trained Loop-OPF can adapt to unseen contingencies of line or generator outages at inference (Section~\ref{sec:topology}).
\item \textbf{Memory-efficient learning of model parameters.} %
{\mohamad Using self-supervised  learning, without OPF labels, Loop-OPF learns the best model parameters across OPF scenarios via  DEQ-based training that keeps the memory fixed}%
(Sections~\ref{sec:training} and~\ref{sec:results_solver}). 
\end{enumerate}

The remainder of this paper is organized as follows. Section~\ref{sec:problem} formulates the OPF problem and introduces the PDHG solver. Section~\ref{sec:method} presents Loop-OPF, together with its optimality guarantees and topology adaptivity, and Section~\ref{sec:training} discusses its memory-efficient training. Section~\ref{sec:results} reports the numerical tests, and Section~\ref{sec:conclusion} concludes the paper.

\section{OPF Formulation and the PDHG Solver}
\label{sec:problem}
\label{sec:opf}

 We formulate the generic OPF problem and introduce the powerful PDHG solver, which will be used later to design our solver-informed Loop-OPF. For simplicity, this paper focuses on the linearized DC-OPF. %
 However, both the formulation and the PDHG solver, and thus our proposed Loop-OPF, can be generalized to include security constraints or AC power flow (PF) models, as discussed later in Remark~\ref{rem:general}.  Linearized PF models around given operating points, notably the DC-PF~\cite{stott2009dcpf}, have been widely used for grid optimization tasks due to the complexity of non-convex AC-OPF~\cite{lehmann2016acnphard}. Moreover, focusing on DC-OPF can facilitate comparisons with existing work, as it has been extensively benchmarked in the learning-for-OPF  literature~\cite{pan2019deepopf, zhao2020deepopfplus, deka2019activesets, park2024pdlscopf}. 

We consider a power network with $N$ buses and $L$ branches. Let $\bbp\in\R^{N}$ and $\bbd\in\R^{N}$ respectively denote the active power generated and consumed at all buses. With the demand $\bbd$ and the dispatch limits  $[\underline{\bbp},\bar{\bbp}]$ given, the DC-OPF seeks to determine the optimal dispatch $\bbp^{\star}$ by solving
\begin{subequations}
\label{eq:dcopf}
\begin{align}
\min_{\underline{\bbp}\,\le\,\bbp\,\le\,\bar{\bbp}} \quad & f(\bbp):= \tfrac{1}{2} \bbp^\top \bbQ \bbp + \bbq^\top \bbp \label{eq:dcopf_obj}\\
\text{s.t.} \quad & \mathbf{1}^\top\bbp = \mathbf{1}^\top\bbd, \label{eq:dcopf_bal}\\
& \bbH\bbp \le \bbg(\bbd).  \label{eq:dcopf_flow}
\end{align}
\end{subequations}
Here, we use a quadratic cost function~\eqref{eq:dcopf_obj}  specified by the diagonal matrix $\bbQ\succeq 0$ and linear coefficients $\bbq$. The cost function can be generalized  to include (piecewise) linear ones, as discussed later. To account for buses with no local generation, one can simply set
$\underline{p}_i=\bar{p}_i=0$. Regarding linear constraints, \eqref{eq:dcopf_bal} ensures the balance of power of the network, while \eqref{eq:dcopf_flow} enforces the thermal limits of all branches. Specifically, the DC-PF model forms the line flow vector $\bbf = \bbS(\bbp-\bbd)\in\R^{L}$ using the injection shift factor (ISF) matrix $\bbS\in\R^{L\times N}${\mohamad~\cite{wood2014power}}. To satisfy the thermal limits given by
$\bbf\in[-\bar{\bbf},\bar{\bbf}]$, we have 
\begin{equation}
\begin{bmatrix} \bbS \\ -\bbS \end{bmatrix} \bbp
\;\le\;
\begin{bmatrix} \bar{\bbf} + \bbS\bbd \\ \bar{\bbf} - \bbS\bbd \end{bmatrix},
\label{eq:doubled_flow}
\end{equation}
which defines $\bbH\in\R^{2L\times N}$ and $\bbg(\bbd)\in\R^{2L}$
in~\eqref{eq:dcopf_flow}. {\mohamad We refer to the network data $(\bbH,\bbg(\bbd),\underline{\bbp},\bar{\bbp})$ in~\eqref{eq:dcopf} as an OPF \textit{scenario}. A new scenario $c$, such as a line outage that changes the topology or a generator outage, alters only these data, denoted with the superscript $(c)$, as detailed in Remark~\ref{rem:general}.} Note that the active line flow constraints, or the network congestion patterns, tend to vary drastically within daily operations, making them very challenging to predict by learning-for-OPF methods.   

\subsection{Primal--Dual Hybrid Gradient (PDHG) Method for OPF}
\label{sec:pdhg}

Our solver-informed Loop-OPF builds on the primal--dual hybrid gradient (PDHG) algorithm~\cite{chambolle2011pdhg}. Owing to its excellent capability in parallel implementation, PDHG is becoming very popular for large-scale optimization tasks including grid optimization; see e.g.,~\cite{sharadga2025gpuuc, sharadga2025scopf}{\mohamad,~\cite{nguyen2026scalable}}. Its iterative updates mainly use
matrix--vector products and element-wise operations, unlike the classical interior-point methods that  require iteratively solving coupled linear systems~\cite{torres1998ipm}. %
PDHG also differs from other well-known iterative methods that need to update their variables sequentially, such as the alternating direction method of multipliers (ADMM)~\cite{boyd2011admm, applegate2021pdlp}. By eliminating these types of updates, each PDHG iteration can be executed inexpensively in parallel, making it appealing for large-scale optimization and also our solver-informed OPF design.

One PDHG iteration consists of both primal and dual updates. The former includes both the primal variable $\bbp$ in \eqref{eq:dcopf} and an auxiliary variable $\bbv$. The dual variables are  the power-balance
multiplier $\lambda$ for the  equality in~\eqref{eq:dcopf_bal} and the non-negative multipliers $\bbnu\geq \mathbf{0}$ for the $2L$ inequalities in~\eqref{eq:dcopf_flow}. These updates use all positive step-sizes, but they can differ across entries: a scalar $\sigma_{\lambda}$ for  $\lambda$,  $\bbsigma\in\R^{2L}$ for $\bbnu$, while 
$\bbtau\in\R^{N}$ for the primal $\bbv$ updates. By defining two diagonal matrices of $\bbD_{\bbtau}=\diag(\bbtau)$ and
$\bbD_{\bbsigma}=\diag(\bbsigma)$, we can express one PDHG iteration as
\begin{subequations}
\label{eq:pdhg}
\begin{empheq}[left={\text{dual}\empheqlbrace\;}]{align}
\lambda^{k+1} &= \lambda^{k} + \sigma_{\lambda}\bigl(\mathbf{1}^\top\bbp^k - \mathbf{1}^\top\bbd\bigr), \label{eq:meth_lambda}\\
\bbnu^{k+1}   &= \bigl[\bbnu^{k} + \bbD_{\bbsigma}\bigl(\bbH\bbp^k - \bbg(\bbd)\bigr)\bigr]_+, \label{eq:meth_nu}
\end{empheq}
\begin{empheq}[left={\text{primal}\empheqlbrace\;}]{align}
\bbv^{k+1}    &= \bbp^k - \bbD_{\bbtau}\bigl(\lambda^{k+1}\mathbf{1} + \bbH^\top\bbnu^{k+1}\bigr), \label{eq:meth_v}\\
\bbp^{k+1}    &= \Pi_{[\underline{\bbp},\bar{\bbp}]}\bigl((\bbI{+}\bbD_{\bbtau}\bbQ)^{-1}(\bbv^{k+1}{-}\bbD_{\bbtau}\bbq)\bigr) \label{eq:meth_p}
\end{empheq}
\end{subequations}
with projections $[\cdot]_+$ onto the positive orthant and $\Pi_{[\underline{\bbp},\bar{\bbp}]}$ onto dispatch limits, respectively. The primal update~\eqref{eq:meth_p} is based on  applying the so-termed \textit{proximal
operator} of the objective function $f$, given by  
\begin{align*}
\bbp^{k+1}&= \Pi_{[\underline{\bbp},\bar{\bbp}]} \big[\prox^{f}_{\bbD_{\bbtau}}(\bbv^{k+1})\big]~~~\mathrm{with}\\
\prox^{f}_{\bbD_{\bbtau}}(\bbv) &:= \arg\min_{\bbx}\, f(\bbx) +
\tfrac{1}{2}(\bbx-\bbv)^\top\bbD_{\bbtau}^{-1}(\bbx-\bbv).
\end{align*}
For the quadratic $f$ in~\eqref{eq:dcopf} with both $\bbQ\succeq \mathbf 0$ and $\bbD_{\bbtau}\succ \mathbf 0$,  the resultant Hessian matrix $(\bbI + \bbD_{\bbtau}\bbQ) \succ \mathbf 0$ is always invertible, and thus this proximal step enjoys a closed-form solution as shown in \eqref{eq:meth_p}.
Note that this proximal step also offers the main improvement in PDHG over  basic primal--dual
gradient methods, since the additional quadratic penalty therein
enhances the numerical stability of primal updates and eliminates the
requirement of differentiable objective functions~\cite{parikh2014prox}. This key advantage facilitates the PDHG solver to incorporate non-differential piecewise linear cost functions or general AC-OPF formulation, as summarized in the following remark.

\begin{remark}[Generalizability]
\label{rem:general}
The problem formulated in~\eqref{eq:dcopf} and the proposed PDHG updates in~\eqref{eq:pdhg} can be generalized to include practical considerations.
First, the cost $f(\bbp)$ can be any convex function, and one practical choice is the \textit{piecewise linear function}~\cite{zimmerman2011matpower,
frangioni2009milpuc, coffrin2021pwlcost}. While we focus on the quadratic form for mathematical convenience, the ensuing results for our proposed Loop-OPF also hold for piecewise linear costs. Second, our approach is also readily applicable for \textit{security-constrained OPF} (SC-OPF)~\cite{velloso2021scdcopf,
park2024pdlscopf}, more commonly used for grid operations. SC-OPF essentially expands~\eqref{eq:doubled_flow} with the linear security constraints for a given contingency set $\mathcal{C}$:
\begin{align}
-\bar{\bbf} \le \bbS^{(c)}(\bbp-\bbd) \le \bar{\bbf},
  \ \ \forall\, c\in \mathcal{C} \label{eq:scopf_cont}
\end{align}
where each ISF matrix $\bbS^{(c)}$ for any line-outage scenario $c$ corresponds to  the post-contingency network{\mohamad, and can be obtained from the base-case ISF matrix via line outage distribution factors~\cite{guo2009lodf}}. {\mohamad Similarly for a scenario $c$ of generator outage at bus $i$, we can update its limits to $\underline{p}^{(c)}_i=\bar{p}^{(c)}_i=0$ with unchanged line flow constraints.} Hence, PDHG and Loop-OPF directly follow after augmenting  $\bbH$ and $\bbg(\bbd)$ in \eqref{eq:dcopf_flow} with \eqref{eq:scopf_cont},  yet with a higher computation cost due to a larger number of constraints. Third, even for the \textit{general AC-OPF}, the PDHG iterations are applicable and it can be shown that any converged solution is  a stationary point for the nonconvex AC-OPF. This is
because the equilibrium of the updates in~\eqref{eq:pdhg} always satisfies the KKT
conditions, as discussed more in Section~\ref{sec:method}. Moreover, the proximal step in~\eqref{eq:meth_p} remains unchanged, since the generation cost $f$ depends only on the active power $\bbp$. Having a convex $f$ could further facilitate the convergence of PDHG, as shown by~\cite{clason2019nonconvexpd}. 
\end{remark}

\section{Solver-Informed Loop-OPF Learning} %
\label{sec:method}

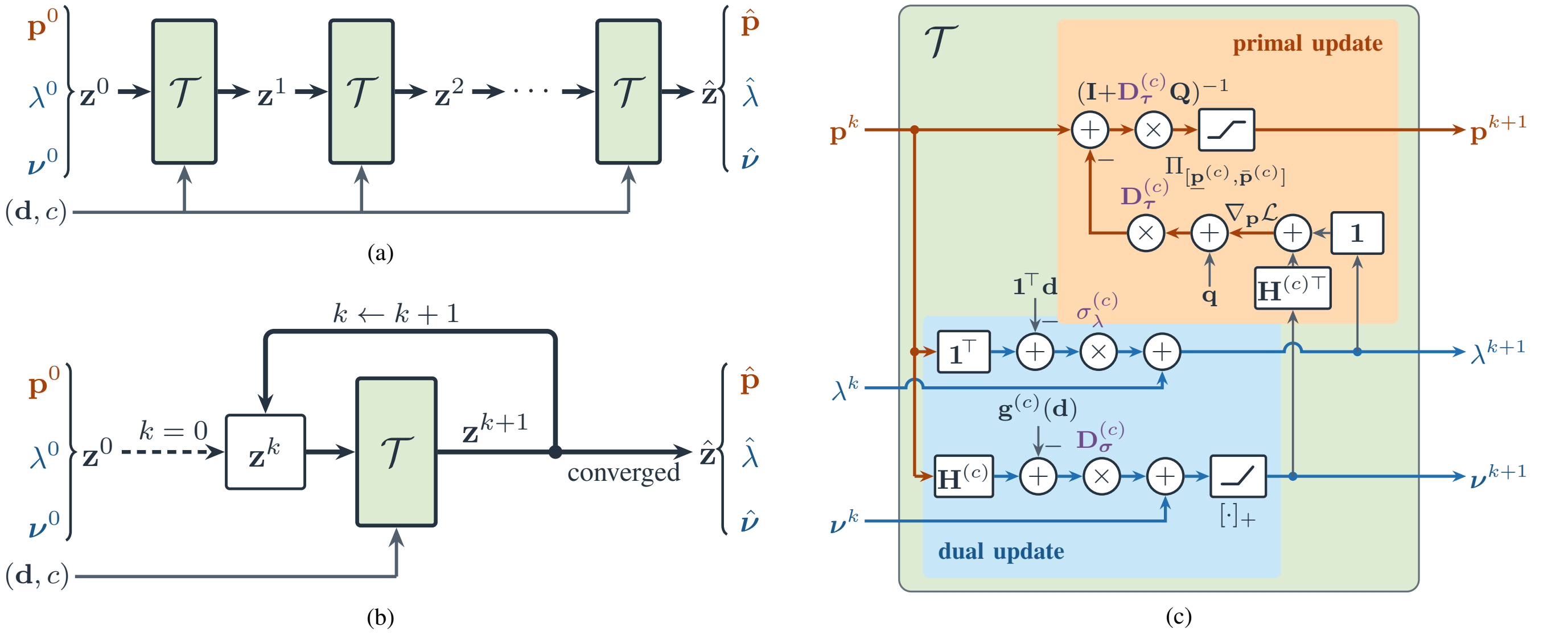
\begin{figure*}[t]
    \centering
    \providecommand{\figCw}{}\renewcommand{\figCw}{0.85\textwidth}%
    \newif\iffigonewide \figonewidefalse %
    \iffigonewide
    \sbox{\figleftbox}{\edef\deqRestoreCatcodes{\catcode32=\the\catcode32\relax\catcode13=\the\catcode13\relax}%
\catcode32=9\relax\catcode13=9\relax
\definecolor{cPrimal}{HTML}{A64209}    %
\definecolor{cDual}{HTML}{1F6FB2}      %
\definecolor{cInk}{HTML}{253341}       %
\definecolor{cGray}{HTML}{52606D}      %
\definecolor{cLayerBg}{HTML}{DDEBD2}   %
\colorlet{cDualText}{cDual!80!black}
\definecolor{cAccent}{HTML}{7A5195}    %
\colorlet{cAccentText}{cAccent!92!black}
\def\fT{\fontsize{12}{14}\selectfont}    %
\def\fbig{\fontsize{10}{12}\selectfont}  %
\def\fvar{\fontsize{9}{11}\selectfont}   %
\def\fnote{\fontsize{8}{10}\selectfont}  %
\def\Tw{6.6mm}\def\Th{15mm}            %
\def\wire{3.5mm}                       %
\def\stackgap{3.8mm}                   %
\def\busdrop{5mm}                      %
\newif\ifshowdata \showdatatrue        %
\def\thetarise{4.2mm}                  %
\newif\ifshowtheta \showthetafalse     %
\deqRestoreCatcodes%
\begin{tikzpicture}[text=cInk, node distance=\wire,
  >={Stealth[length=2mm,width=1.6mm]},
  Tblk/.style={draw=cInk,line width=1.1pt,fill=cLayerBg,rounded corners=1pt,
               minimum width=\Tw,minimum height=\Th,inner sep=0pt,font=\fT},
  zlab/.style={font=\fbig,text=cInk,inner sep=1.6pt},
  lPrimal/.style={font=\fvar,text=cPrimal,inner sep=0.6pt},
  lDual/.style={font=\fvar,text=cDualText,inner sep=0.6pt},
  lConst/.style={font=\fnote,text=cInk,inner sep=1pt},
  wState/.style={cInk,line width=1.4pt},
  wConst/.style={cGray,line width=0.95pt},
  wLearn/.style={cAccent,line width=0.95pt},
  lLearn/.style={font=\fnote,text=cAccentText,inner sep=1pt},
  brc/.style={decorate,decoration={brace,amplitude=2.6pt},cInk,line width=0.75pt}]

\node[lDual] (l0) {$\lambda^{0}$};
\node[lPrimal,above=\stackgap of l0] (p0) {$\mathbf{p}^{0}$};
\node[lDual,below=\stackgap of l0]   (n0) {$\bm{\nu}^{0}$};
\coordinate (b0x) at ($(l0.east)+(0.5pt,0)$);
\draw[brc] ($(b0x |- p0.north)$) -- ($(b0x |- n0.south)$)
  node[midway,right=2.6pt,zlab] (z0) {$\mathbf{z}^{0}$};

\node[Tblk,right=of z0] (T1) {$\mathcal{T}$};
\node[zlab,right=of T1] (z1) {$\mathbf{z}^{1}$};
\node[Tblk,right=of z1] (T2) {$\mathcal{T}$};
\node[zlab,right=of T2] (z2) {$\mathbf{z}^{2}$};
\node[zlab,right=of z2] (dots) {$\cdots$};
\node[Tblk,right=of dots] (TK) {$\mathcal{T}$};
\node[zlab,right=of TK] (zs) {$\hat{\mathbf{z}}$};

\coordinate (bsx) at ($(zs.east)+(1.2pt,0)$);
\node[lDual,anchor=west] (ls) at ($(bsx)+(3.1pt,0)$) {$\hat{\lambda}$};
\node[lPrimal,above=\stackgap of ls] (ps) {$\hat{\mathbf{p}}$};
\node[lDual,below=\stackgap of ls]   (ns) {$\hat{\bm{\nu}}$};
\draw[brc] ($(bsx |- ns.south)$) -- ($(bsx |- ps.north)$);

\foreach \a/\b in {z0/T1, T1/z1, z1/T2, T2/z2, z2/dots, dots/TK, TK/zs}
  \draw[wState,->] (\a) -- (\b);

\ifshowdata
  \coordinate (bus) at ($(T1.south)+(0,-\busdrop)$);
  \node[lConst,anchor=east] (dc) at ($(bus -| z0.west)+(0,0)$) {$(\mathbf{d},c)$};
  \draw[wConst] (dc.east) -- (bus -| TK);
  \foreach \t in {T1,T2,TK} \draw[wConst,->] (bus -| \t) -- (\t.south);
\fi
\ifshowtheta
  \coordinate (tbus) at ($(T1.north)+(0,\thetarise)$);
  \node[lLearn,anchor=east] (th) at (tbus -| z0.west) {$\bm{\theta}$};
  \draw[wLearn] (th.east) -- (tbus -| TK);
  \foreach \t in {T1,T2,TK} \draw[wLearn,->] (tbus -| \t) -- (\t.north);
\fi
\end{tikzpicture}%
}%
    \sbox{\figcolbox}{\edef\deqRestoreCatcodes{\catcode32=\the\catcode32\relax\catcode13=\the\catcode13\relax}%
\catcode32=9\relax\catcode13=9\relax
\definecolor{cPrimal}{HTML}{A64209}    %
\definecolor{cDual}{HTML}{1F6FB2}      %
\definecolor{cInk}{HTML}{253341}       %
\definecolor{cGray}{HTML}{52606D}      %
\definecolor{cLayerBg}{HTML}{DDEBD2}   %
\colorlet{cDualText}{cDual!80!black}
\definecolor{cAccent}{HTML}{7A5195}    %
\colorlet{cAccentText}{cAccent!92!black}
\def\fT{\fontsize{12}{14}\selectfont}    %
\def\fbig{\fontsize{10}{12}\selectfont}  %
\def\fvar{\fontsize{9}{11}\selectfont}   %
\def\fnote{\fontsize{8}{10}\selectfont}  %
\def\Tw{8mm}\def\Th{15mm}              %
\def\Sw{8mm}\def\Sh{7.5mm}             %
\def\initwire{10.5mm}                   %
\def\readwire{5mm}                     %
\def\kpwire{12mm}                      %
\def\exitwire{14mm}                    %
\def\looprise{4.6mm}                   %
\def\stackgap{3.8mm}                   %
\def\busdrop{5mm}                      %
\newif\ifshowdata \showdatatrue        %
\def\feedsep{1.8mm}                    %
\newif\ifshowtheta \showthetafalse     %
\deqRestoreCatcodes%
\begin{tikzpicture}[text=cInk,
  >={Stealth[length=2mm,width=1.6mm]},
  Tblk/.style={draw=cInk,line width=1.1pt,fill=cLayerBg,rounded corners=1pt,
               minimum width=\Tw,minimum height=\Th,inner sep=0pt,font=\fT},
  Sblk/.style={draw=cInk,line width=0.8pt,fill=white,rounded corners=1pt,
               minimum width=\Sw,minimum height=\Sh,inner sep=0pt,font=\fbig},
  zlab/.style={font=\fbig,text=cInk,inner sep=1.6pt},
  lPrimal/.style={font=\fvar,text=cPrimal,inner sep=0.6pt},
  lDual/.style={font=\fvar,text=cDualText,inner sep=0.6pt},
  lConst/.style={font=\fnote,text=cInk,inner sep=1pt},
  lNote/.style={font=\fnote,text=cInk,inner sep=1.2pt},
  wState/.style={cInk,line width=1.4pt},
  wInit/.style={cInk,line width=1.1pt,dash pattern=on 2.6pt off 1.8pt},
  wConst/.style={cGray,line width=0.95pt},
  wLearn/.style={cAccent,line width=0.95pt},
  lLearn/.style={font=\fnote,text=cAccentText,inner sep=1pt},
  dot/.style={circle,fill=cInk,inner sep=0pt,minimum size=4pt},
  brc/.style={decorate,decoration={brace,amplitude=2.6pt},cInk,line width=0.75pt}]

\node[lDual] (l0) {$\lambda^{0}$};
\node[lPrimal,above=\stackgap of l0] (p0) {$\mathbf{p}^{0}$};
\node[lDual,below=\stackgap of l0]   (n0) {$\bm{\nu}^{0}$};
\coordinate (b0x) at ($(l0.east)+(0.5pt,0)$);
\draw[brc] ($(b0x |- p0.north)$) -- ($(b0x |- n0.south)$)
  node[midway,right=2.6pt,zlab] (z0) {$\mathbf{z}^{0}$};

\node[Sblk,anchor=west] (S) at ($(z0.east)+(\initwire,0)$) {$\mathbf{z}^{k}$};
\node[Tblk,anchor=west] (T) at ($(S.east)+(\readwire,0)$) {$\mathcal{T}$};
\coordinate (jout) at ($(T.east)+(\kpwire,0)$);                  %
\node[zlab,anchor=west] (zs) at ($(jout)+(\exitwire,0)$) {$\hat{\mathbf{z}}$};

\coordinate (bsx) at ($(zs.east)+(1.2pt,0)$);
\node[lDual,anchor=west] (ls) at ($(bsx)+(3.1pt,0)$) {$\hat{\lambda}$};
\node[lPrimal,above=\stackgap of ls] (ps) {$\hat{\mathbf{p}}$};
\node[lDual,below=\stackgap of ls]   (ns) {$\hat{\bm{\nu}}$};
\draw[brc] ($(bsx |- ns.south)$) -- ($(bsx |- ps.north)$);

\draw[wInit,->] (z0) -- node[above=2pt,lNote] {$k=0$} (S);
\draw[wState,->] (S) -- (T);
\draw[wState] (T) -- node[above=0.5pt,zlab] {$\mathbf{z}^{k+1}$} (jout);
\coordinate (ltop) at ($(T.north)+(0,\looprise)$);
\draw[wState,->,rounded corners=2.5pt] (jout) |- (ltop) -| (S.north);
\node[lNote,above=0.5pt] at (ltop) {$k\leftarrow k+1$};
\draw[wState,->] (jout) -- node[below=1pt,lNote] {converged} (zs);
\node[dot] at (jout) {};

\coordinate (bus) at ($(T.south)+(0,-\busdrop)$);
\ifshowtheta \coordinate (inDC) at ($(T.south)+(-\feedsep,0)$);
\else        \coordinate (inDC) at (T.south); \fi
\ifshowdata
  \node[lConst,anchor=east] (dc) at ($(bus -| z0.west)$) {$(\mathbf{d},c)$};
  \draw[wConst] (dc.east) -- (bus -| inDC);
  \draw[wConst,->] (bus -| inDC) -- (inDC);
\fi
\ifshowtheta
  \coordinate (inTh) at ($(T.south)+(\feedsep,0)$);
  \node[lLearn,anchor=west] (th) at (bus -| zs.west) {$\bm{\theta}$};
  \draw[wLearn] (th.west) -- (bus -| inTh);
  \draw[wLearn,->] (bus -| inTh) -- (inTh);
\fi
\end{tikzpicture}%
}%
    \subfloat[\label{fig:deq_chain}]{\resizebox{\columnwidth}{!}{\usebox{\figleftbox}}}\hfill
    \subfloat[\label{fig:deq_loop}]{\resizebox{\columnwidth}{!}{\usebox{\figcolbox}}}\par\medskip
    \sbox{\figleftbox}{\edef\pdhgRestoreCatcodes{\catcode32=\the\catcode32\relax\catcode13=\the\catcode13\relax}%
\catcode32=9\relax\catcode13=9\relax
\definecolor{cPrimal}{HTML}{A64209}    %
\definecolor{cDual}{HTML}{1F6FB2}      %
\definecolor{cAccent}{HTML}{7A5195}    %
\definecolor{cInk}{HTML}{253341}       %
\definecolor{cGray}{HTML}{52606D}      %
\definecolor{cLayerBg}{HTML}{DDEBD2}   %
\definecolor{cPrimalBg}{HTML}{FFD9B0}  %
\definecolor{cDualBg}{HTML}{C7E6F7}    %
\colorlet{cDualText}{cDual!80!black}   %
\colorlet{cAccentText}{cAccent!92!black}
\def\fvar{\fontsize{9}{11}\selectfont}   %
\def\fT{\fontsize{14}{17}\selectfont}   %
\def\fmain{\fontsize{9}{11}\selectfont}  %
\def\fhead{\fontsize{8}{10}\selectfont}  %
\def\fnote{\fontsize{8}{10}\selectfont}  %
\def\opsize{4.2mm}                     %
\def\matw{6mm}\def\matht{4.8mm}        %
\def\actw{6.4mm}                       %
\def\gap{2.9mm}                        %
\def\gradlen{5.2mm}                    %
\def\rowgap{9.4mm}                     %
\def\jn{3mm}                           %
\def\jsep{7.5mm}                       %
\def\HTrise{7.4mm}                     %
\def\Grise{13.8mm}                     %
\def\Prise{12mm}                       %
\def\svback{6.5mm}                     %
\def\fanlen{2.6mm}                     %
\def\frameL{1.8mm}\def\frameR{2.4mm}   %
\def\frameT{1.6mm}\def\frameB{1.6mm}   %
\def\inside{4mm}\def\outside{5.5mm}    %
\def\skipL{4.2mm}\def\skipN{5.2mm}     %
\def\hop{1mm}                          %
\def\biasL{4mm}\def\biasN{3.6mm}\def\biasq{4mm} %
\pdhgRestoreCatcodes%
\begin{tikzpicture}[font=\fmain, text=cInk, node distance=\gap,
  >={Stealth[length=1.6mm,width=1.3mm]},
  mat/.style={draw=cInk,line width=0.8pt,fill=white,rounded corners=0.8pt,
              minimum width=\matw,minimum height=\matht,inner sep=0.6pt,text=cInk},
  op/.style={circle,draw=cInk,line width=0.8pt,fill=white,text=cInk,
             minimum size=\opsize,inner sep=0pt},
  actb/.style={draw=cInk,line width=0.8pt,fill=white,rounded corners=0.8pt,
               minimum width=\actw,minimum height=\matht,inner sep=0pt},
  glyph/.style={cInk,line width=0.9pt,line cap=round,line join=round},
  wPrimal/.style={cPrimal,line width=1.0pt},   %
  wDual/.style={cDual,line width=1.0pt},       %
  wShared/.style={cGray,line width=0.7pt},    %
  wConst/.style={cGray,line width=0.7pt},     %
  lPrimal/.style={font=\fvar,text=cPrimal,inner sep=1pt},
  lDual/.style={font=\fvar,text=cDualText,inner sep=1pt},
  lConst/.style={font=\fnote,text=cInk,inner sep=1pt},
  lLearn/.style={font=\fnote,text=cAccentText,inner sep=1pt}, %
  lAnno/.style={font=\fnote,text=cInk,inner sep=1pt},
  lSign/.style={font=\fnote,text=cInk,inner sep=0.3pt},
  hPrimal/.style={font=\fhead\bfseries,text=cPrimal},
  hDual/.style={font=\fhead\bfseries,text=cDualText},
  region/.style={rounded corners=2pt,inner sep=1.6mm,fill=#1},
  tframe/.style={draw=cInk!70,line width=0.6pt,fill=cLayerBg,rounded corners=4pt}, %
  lT/.style={font=\fT,text=cInk},
  dot/.style={circle,fill=#1,inner sep=0pt,minimum size=3pt}]

\node[mat] (oneT) {$\mathbf{1}^{\!\top}$};
\node[op,right=of oneT] (sa) {$+$};
\node[op,right=of sa]   (ma) {$\times$};
\node[op,right=of ma]   (sb) {$+$};
\node[lLearn,above=0.5pt of ma] {$\sigma_{\lambda}^{(c)}$};
\node[mat,below=\rowgap of oneT] (Hm) {$\mathbf{H}^{(c)}$};
\node[op,right=of Hm] (sc) {$+$};
\node[op,right=of sc] (mc) {$\times$};
\node[op,right=of mc] (sd) {$+$};
\node[lLearn,above=0.5pt of mc] {$\mathbf{D}_{\bm{\sigma}}^{(c)}$};
\node[actb,right=of sd] (relu) {};
\draw[glyph] ($(relu)+(-0.19,-0.09)$)--($(relu)+(0,-0.09)$)--($(relu)+(0.17,0.11)$);
\node[lAnno,below=0.5pt of relu] (relulab) {$[\cdot]_+$};

\coordinate (jn) at ($(relu.east)+(\jn,0)$);          %
\coordinate (jl) at ($(jn |- sb)+(\jsep,0)$);          %

\node[mat] (one) at ($(jl)+(0,\Grise)$) {$\mathbf{1}$};
\node[op]  (sg)  at (jn |- one) {$+$};
\node[mat] (HT)  at ($(jn |- sb)+(0,\HTrise)$) {$\mathbf{H}^{(c)\top}$};
\node[op,left=\gradlen of sg] (sq) {$+$};
\node[op,left=of sq] (mg) {$\times$};
\node[lLearn,above=0.5pt of mg] {$\mathbf{D}_{\bm{\tau}}^{(c)}$};
\node[op] (sv) at ($(mg)+(-\svback,\Prise)$) {$+$};
\node[op,right=of sv] (mq) {$\times$};
\node[lConst,above=0.5pt of mq] (mqlab) {$(\mathbf{I}{+}\textcolor{cAccentText}{\mathbf{D}_{\bm{\tau}}^{(c)}}\mathbf{Q})^{-1}$}; %
\node[actb,right=of mq] (clip) {};
\draw[glyph] ($(clip)+(-0.21,-0.09)$)--($(clip)+(-0.08,-0.09)$)--($(clip)+(0.08,0.09)$)--($(clip)+(0.21,0.09)$);
\node[lAnno,below=0.5pt of clip] (cliplab) {$\Pi_{[\underline{\mathbf{p}}^{(c)},\bar{\mathbf{p}}^{(c)}]}$};

\coordinate (fan) at ($(oneT.west)+(-\fanlen,0)$);
\coordinate (inP) at ($(fan |- sv)+(-\frameL-\inside,0)$);
\coordinate (inL) at ($(inP |- sb)+(0,-\skipL)$);
\coordinate (inN) at ($(inP |- sd)+(0,-\skipN)$);
\coordinate (outx) at ($(one.east)+(1.6mm+\frameR+\outside,0)$);  %
\node[left=0pt of inP,lPrimal] (lblPk) {$\mathbf{p}^{k}$};
\node[left=0pt of inL,lDual]   (lblLk) {$\lambda^{k}$};
\node[left=0pt of inN,lDual]   (lblNk) {$\bm{\nu}^{k}$};
\node[right=0pt,lPrimal] (lblPo) at (outx |- sv) {$\mathbf{p}^{k+1}$};
\node[right=0pt,lDual]   (lblLo) at (outx |- sb) {$\lambda^{k+1}$};
\node[right=0pt,lDual]   (lblNo) at (outx |- sd) {$\bm{\nu}^{k+1}$};

\coordinate (dbot) at ($(oneT.west |- inN)+(0,-0.5)$);
\coordinate (ptop) at ($(sv.west |- mqlab.north)+(0,0.42)$);
\coordinate (frTL) at ($(fan |- ptop)+(-\frameL,1.6mm+\frameT)$);       %
\coordinate (frBR) at ($(one.east |- dbot)+(1.6mm+\frameR,-1.6mm-\frameB)$);
\begin{scope}[on background layer]
  \draw[tframe] (frTL) rectangle (frBR);
  \node[region=cDualBg,fit=(oneT)(sa)(relu)(relulab)(dbot)] (dbox) {};
  \node[region=cPrimalBg,fit=(sv)(clip)(cliplab)(mqlab)(one)(HT)(mg)(ptop)] (pbox) {};
\end{scope}
\node[hDual,anchor=south west]   at ($(dbox.south west)+(0.06,0.05)$) {dual update};
\node[hPrimal,anchor=north east] at ($(pbox.north east)+(-0.06,-0.05)$) {primal update};
\node[lT,anchor=north west] at ($(frTL)+(1.6mm,-1.2mm)$) {$\mathcal{T}$};

\draw[wDual,->] (inL) -- ($(fan |- inL)+(-\hop,0)$) arc[start angle=180,end angle=0,radius=\hop] -| (sb);  %
\draw[wDual,->] (inN) -| (sd);
\draw[wPrimal,->] (inP) -- (sv);
\node[dot=cPrimal] at (fan |- sv) {};
\draw[wPrimal] (fan |- sv) -- (fan |- Hm);
\node[dot=cPrimal] at (fan) {};
\draw[wPrimal,->] (fan) -- (oneT);
\draw[wPrimal,->] (fan |- Hm) -- (Hm);
\draw[wDual,->] (oneT) -- (sa);
\draw[wDual,->] (sa) -- (ma);
\draw[wDual,->] (ma) -- (sb);
\draw[wDual,->] (Hm) -- (sc);
\draw[wDual,->] (sc) -- (mc);
\draw[wDual,->] (mc) -- (sd);
\draw[wDual,->] (sd) -- (relu);
\draw[wConst,->] ($(sa.north)+(0,\biasL)$) node[above,lConst] {$\mathbf{1}^{\!\top}\mathbf{d}$} -- (sa);
\node[lSign,anchor=south west] at ($(sa.north)+(0.04,0.01)$) {$-$};
\draw[wConst,->] ($(sc.north)+(0,\biasN)$) node[above,lConst] {$\mathbf{g}^{(c)}(\mathbf{d})$} -- (sc);
\node[lSign,anchor=south west] at ($(sc.north)+(0.04,0.01)$) {$-$};
\draw[wConst,->] ($(sq.south)+(0,-\biasq)$) node[below,lConst] {$\mathbf{q}$} -- (sq);  %
\draw[wDual,->] (sb) -- ($(jn |- sb)+(-\hop,0)$) arc[start angle=180,end angle=0,radius=\hop] -- (outx |- sb);
\node[dot=cDual] at (jl) {};
\draw[wShared,->] (jl) -- (one);
\draw[wDual,->] (relu) -- (outx |- sd);
\node[dot=cDual] at (jn) {};
\draw[wShared,->] (jn) -- (HT);
\draw[wShared,->] (one) -- (sg);
\draw[wShared,->] (HT) -- (sg);
\draw[wPrimal,->] (sg) -- node[above=0.5pt,lConst] {$\nabla_{\mathbf{p}}\mathcal{L}$} (sq);
\draw[wPrimal,->] (sq) -- (mg);
\draw[wPrimal,->] (mg) -| (sv);
\node[lSign,anchor=north west] at ($(sv.south)+(0.05,-0.01)$) {$-$};
\draw[wPrimal,->] (sv) -- (mq);
\draw[wPrimal,->] (mq) -- (clip);
\draw[wPrimal,->] (clip) -- (outx |- sv);
\end{tikzpicture}%
}%
    \subfloat[\label{fig:pdhg_layer}]{\resizebox{\figCw}{!}{\usebox{\figleftbox}}}%
    \else
    \setcounter{subfigure}{2}\setcounter{subfigure@save}{2}%
    \sbox0{\edef\pdhgRestoreCatcodes{\catcode32=\the\catcode32\relax\catcode13=\the\catcode13\relax}%
\catcode32=9\relax\catcode13=9\relax
\definecolor{cPrimal}{HTML}{A64209}    %
\definecolor{cDual}{HTML}{1F6FB2}      %
\definecolor{cAccent}{HTML}{7A5195}    %
\definecolor{cInk}{HTML}{253341}       %
\definecolor{cGray}{HTML}{52606D}      %
\definecolor{cLayerBg}{HTML}{DDEBD2}   %
\definecolor{cPrimalBg}{HTML}{FFD9B0}  %
\definecolor{cDualBg}{HTML}{C7E6F7}    %
\colorlet{cDualText}{cDual!80!black}   %
\colorlet{cAccentText}{cAccent!92!black}
\def\fvar{\fontsize{9}{11}\selectfont}   %
\def\fT{\fontsize{14}{17}\selectfont}   %
\def\fmain{\fontsize{9}{11}\selectfont}  %
\def\fhead{\fontsize{8}{10}\selectfont}  %
\def\fnote{\fontsize{8}{10}\selectfont}  %
\def\opsize{4.2mm}                     %
\def\matw{6mm}\def\matht{4.8mm}        %
\def\actw{6.4mm}                       %
\def\gap{2.9mm}                        %
\def\gradlen{5.2mm}                    %
\def\rowgap{9.4mm}                     %
\def\jn{3mm}                           %
\def\jsep{7.5mm}                       %
\def\HTrise{7.4mm}                     %
\def\Grise{13.8mm}                     %
\def\Prise{12mm}                       %
\def\svback{6.5mm}                     %
\def\fanlen{2.6mm}                     %
\def\frameL{1.8mm}\def\frameR{2.4mm}   %
\def\frameT{1.6mm}\def\frameB{1.6mm}   %
\def\inside{4mm}\def\outside{5.5mm}    %
\def\skipL{4.2mm}\def\skipN{5.2mm}     %
\def\hop{1mm}                          %
\def\biasL{4mm}\def\biasN{3.6mm}\def\biasq{4mm} %
\pdhgRestoreCatcodes%
\begin{tikzpicture}[font=\fmain, text=cInk, node distance=\gap,
  >={Stealth[length=1.6mm,width=1.3mm]},
  mat/.style={draw=cInk,line width=0.8pt,fill=white,rounded corners=0.8pt,
              minimum width=\matw,minimum height=\matht,inner sep=0.6pt,text=cInk},
  op/.style={circle,draw=cInk,line width=0.8pt,fill=white,text=cInk,
             minimum size=\opsize,inner sep=0pt},
  actb/.style={draw=cInk,line width=0.8pt,fill=white,rounded corners=0.8pt,
               minimum width=\actw,minimum height=\matht,inner sep=0pt},
  glyph/.style={cInk,line width=0.9pt,line cap=round,line join=round},
  wPrimal/.style={cPrimal,line width=1.0pt},   %
  wDual/.style={cDual,line width=1.0pt},       %
  wShared/.style={cGray,line width=0.7pt},    %
  wConst/.style={cGray,line width=0.7pt},     %
  lPrimal/.style={font=\fvar,text=cPrimal,inner sep=1pt},
  lDual/.style={font=\fvar,text=cDualText,inner sep=1pt},
  lConst/.style={font=\fnote,text=cInk,inner sep=1pt},
  lLearn/.style={font=\fnote,text=cAccentText,inner sep=1pt}, %
  lAnno/.style={font=\fnote,text=cInk,inner sep=1pt},
  lSign/.style={font=\fnote,text=cInk,inner sep=0.3pt},
  hPrimal/.style={font=\fhead\bfseries,text=cPrimal},
  hDual/.style={font=\fhead\bfseries,text=cDualText},
  region/.style={rounded corners=2pt,inner sep=1.6mm,fill=#1},
  tframe/.style={draw=cInk!70,line width=0.6pt,fill=cLayerBg,rounded corners=4pt}, %
  lT/.style={font=\fT,text=cInk},
  dot/.style={circle,fill=#1,inner sep=0pt,minimum size=3pt}]

\node[mat] (oneT) {$\mathbf{1}^{\!\top}$};
\node[op,right=of oneT] (sa) {$+$};
\node[op,right=of sa]   (ma) {$\times$};
\node[op,right=of ma]   (sb) {$+$};
\node[lLearn,above=0.5pt of ma] {$\sigma_{\lambda}^{(c)}$};
\node[mat,below=\rowgap of oneT] (Hm) {$\mathbf{H}^{(c)}$};
\node[op,right=of Hm] (sc) {$+$};
\node[op,right=of sc] (mc) {$\times$};
\node[op,right=of mc] (sd) {$+$};
\node[lLearn,above=0.5pt of mc] {$\mathbf{D}_{\bm{\sigma}}^{(c)}$};
\node[actb,right=of sd] (relu) {};
\draw[glyph] ($(relu)+(-0.19,-0.09)$)--($(relu)+(0,-0.09)$)--($(relu)+(0.17,0.11)$);
\node[lAnno,below=0.5pt of relu] (relulab) {$[\cdot]_+$};

\coordinate (jn) at ($(relu.east)+(\jn,0)$);          %
\coordinate (jl) at ($(jn |- sb)+(\jsep,0)$);          %

\node[mat] (one) at ($(jl)+(0,\Grise)$) {$\mathbf{1}$};
\node[op]  (sg)  at (jn |- one) {$+$};
\node[mat] (HT)  at ($(jn |- sb)+(0,\HTrise)$) {$\mathbf{H}^{(c)\top}$};
\node[op,left=\gradlen of sg] (sq) {$+$};
\node[op,left=of sq] (mg) {$\times$};
\node[lLearn,above=0.5pt of mg] {$\mathbf{D}_{\bm{\tau}}^{(c)}$};
\node[op] (sv) at ($(mg)+(-\svback,\Prise)$) {$+$};
\node[op,right=of sv] (mq) {$\times$};
\node[lConst,above=0.5pt of mq] (mqlab) {$(\mathbf{I}{+}\textcolor{cAccentText}{\mathbf{D}_{\bm{\tau}}^{(c)}}\mathbf{Q})^{-1}$}; %
\node[actb,right=of mq] (clip) {};
\draw[glyph] ($(clip)+(-0.21,-0.09)$)--($(clip)+(-0.08,-0.09)$)--($(clip)+(0.08,0.09)$)--($(clip)+(0.21,0.09)$);
\node[lAnno,below=0.5pt of clip] (cliplab) {$\Pi_{[\underline{\mathbf{p}}^{(c)},\bar{\mathbf{p}}^{(c)}]}$};

\coordinate (fan) at ($(oneT.west)+(-\fanlen,0)$);
\coordinate (inP) at ($(fan |- sv)+(-\frameL-\inside,0)$);
\coordinate (inL) at ($(inP |- sb)+(0,-\skipL)$);
\coordinate (inN) at ($(inP |- sd)+(0,-\skipN)$);
\coordinate (outx) at ($(one.east)+(1.6mm+\frameR+\outside,0)$);  %
\node[left=0pt of inP,lPrimal] (lblPk) {$\mathbf{p}^{k}$};
\node[left=0pt of inL,lDual]   (lblLk) {$\lambda^{k}$};
\node[left=0pt of inN,lDual]   (lblNk) {$\bm{\nu}^{k}$};
\node[right=0pt,lPrimal] (lblPo) at (outx |- sv) {$\mathbf{p}^{k+1}$};
\node[right=0pt,lDual]   (lblLo) at (outx |- sb) {$\lambda^{k+1}$};
\node[right=0pt,lDual]   (lblNo) at (outx |- sd) {$\bm{\nu}^{k+1}$};

\coordinate (dbot) at ($(oneT.west |- inN)+(0,-0.5)$);
\coordinate (ptop) at ($(sv.west |- mqlab.north)+(0,0.42)$);
\coordinate (frTL) at ($(fan |- ptop)+(-\frameL,1.6mm+\frameT)$);       %
\coordinate (frBR) at ($(one.east |- dbot)+(1.6mm+\frameR,-1.6mm-\frameB)$);
\begin{scope}[on background layer]
  \draw[tframe] (frTL) rectangle (frBR);
  \node[region=cDualBg,fit=(oneT)(sa)(relu)(relulab)(dbot)] (dbox) {};
  \node[region=cPrimalBg,fit=(sv)(clip)(cliplab)(mqlab)(one)(HT)(mg)(ptop)] (pbox) {};
\end{scope}
\node[hDual,anchor=south west]   at ($(dbox.south west)+(0.06,0.05)$) {dual update};
\node[hPrimal,anchor=north east] at ($(pbox.north east)+(-0.06,-0.05)$) {primal update};
\node[lT,anchor=north west] at ($(frTL)+(1.6mm,-1.2mm)$) {$\mathcal{T}$};

\draw[wDual,->] (inL) -- ($(fan |- inL)+(-\hop,0)$) arc[start angle=180,end angle=0,radius=\hop] -| (sb);  %
\draw[wDual,->] (inN) -| (sd);
\draw[wPrimal,->] (inP) -- (sv);
\node[dot=cPrimal] at (fan |- sv) {};
\draw[wPrimal] (fan |- sv) -- (fan |- Hm);
\node[dot=cPrimal] at (fan) {};
\draw[wPrimal,->] (fan) -- (oneT);
\draw[wPrimal,->] (fan |- Hm) -- (Hm);
\draw[wDual,->] (oneT) -- (sa);
\draw[wDual,->] (sa) -- (ma);
\draw[wDual,->] (ma) -- (sb);
\draw[wDual,->] (Hm) -- (sc);
\draw[wDual,->] (sc) -- (mc);
\draw[wDual,->] (mc) -- (sd);
\draw[wDual,->] (sd) -- (relu);
\draw[wConst,->] ($(sa.north)+(0,\biasL)$) node[above,lConst] {$\mathbf{1}^{\!\top}\mathbf{d}$} -- (sa);
\node[lSign,anchor=south west] at ($(sa.north)+(0.04,0.01)$) {$-$};
\draw[wConst,->] ($(sc.north)+(0,\biasN)$) node[above,lConst] {$\mathbf{g}^{(c)}(\mathbf{d})$} -- (sc);
\node[lSign,anchor=south west] at ($(sc.north)+(0.04,0.01)$) {$-$};
\draw[wConst,->] ($(sq.south)+(0,-\biasq)$) node[below,lConst] {$\mathbf{q}$} -- (sq);  %
\draw[wDual,->] (sb) -- ($(jn |- sb)+(-\hop,0)$) arc[start angle=180,end angle=0,radius=\hop] -- (outx |- sb);
\node[dot=cDual] at (jl) {};
\draw[wShared,->] (jl) -- (one);
\draw[wDual,->] (relu) -- (outx |- sd);
\node[dot=cDual] at (jn) {};
\draw[wShared,->] (jn) -- (HT);
\draw[wShared,->] (one) -- (sg);
\draw[wShared,->] (HT) -- (sg);
\draw[wPrimal,->] (sg) -- node[above=0.5pt,lConst] {$\nabla_{\mathbf{p}}\mathcal{L}$} (sq);
\draw[wPrimal,->] (sq) -- (mg);
\draw[wPrimal,->] (mg) -| (sv);
\node[lSign,anchor=north west] at ($(sv.south)+(0.05,-0.01)$) {$-$};
\draw[wPrimal,->] (sv) -- (mq);
\draw[wPrimal,->] (mq) -- (clip);
\draw[wPrimal,->] (clip) -- (outx |- sv);
\end{tikzpicture}%
}%
    \sbox{\figcolbox}{\parbox[b]{\columnwidth}{\centering
        \subfloat[\label{fig:pdhg_layer}]{\ifdim\wd0>\columnwidth\resizebox{\columnwidth}{!}{\usebox0}\else\usebox0\fi}}}%
    \edef\figcoldp{\the\dp\figcolbox}%
    \sbox{\figcolbox}{\raisebox{\dp\figcolbox}{\usebox{\figcolbox}}}%
    \setlength{\figcolht}{\ht\figcolbox}%
    \setcounter{subfigure}{0}\setcounter{subfigure@save}{0}%
    \sbox{\figleftbox}{\parbox[b][\dimexpr\figcolht-\figcoldp\relax][s]{\columnwidth}{\centering
        \sbox0{\edef\deqRestoreCatcodes{\catcode32=\the\catcode32\relax\catcode13=\the\catcode13\relax}%
\catcode32=9\relax\catcode13=9\relax
\definecolor{cPrimal}{HTML}{A64209}    %
\definecolor{cDual}{HTML}{1F6FB2}      %
\definecolor{cInk}{HTML}{253341}       %
\definecolor{cGray}{HTML}{52606D}      %
\definecolor{cLayerBg}{HTML}{DDEBD2}   %
\colorlet{cDualText}{cDual!80!black}
\definecolor{cAccent}{HTML}{7A5195}    %
\colorlet{cAccentText}{cAccent!92!black}
\def\fT{\fontsize{12}{14}\selectfont}    %
\def\fbig{\fontsize{10}{12}\selectfont}  %
\def\fvar{\fontsize{9}{11}\selectfont}   %
\def\fnote{\fontsize{8}{10}\selectfont}  %
\def\Tw{6.6mm}\def\Th{15mm}            %
\def\wire{3.5mm}                       %
\def\stackgap{3.8mm}                   %
\def\busdrop{5mm}                      %
\newif\ifshowdata \showdatatrue        %
\def\thetarise{4.2mm}                  %
\newif\ifshowtheta \showthetafalse     %
\deqRestoreCatcodes%
\begin{tikzpicture}[text=cInk, node distance=\wire,
  >={Stealth[length=2mm,width=1.6mm]},
  Tblk/.style={draw=cInk,line width=1.1pt,fill=cLayerBg,rounded corners=1pt,
               minimum width=\Tw,minimum height=\Th,inner sep=0pt,font=\fT},
  zlab/.style={font=\fbig,text=cInk,inner sep=1.6pt},
  lPrimal/.style={font=\fvar,text=cPrimal,inner sep=0.6pt},
  lDual/.style={font=\fvar,text=cDualText,inner sep=0.6pt},
  lConst/.style={font=\fnote,text=cInk,inner sep=1pt},
  wState/.style={cInk,line width=1.4pt},
  wConst/.style={cGray,line width=0.95pt},
  wLearn/.style={cAccent,line width=0.95pt},
  lLearn/.style={font=\fnote,text=cAccentText,inner sep=1pt},
  brc/.style={decorate,decoration={brace,amplitude=2.6pt},cInk,line width=0.75pt}]

\node[lDual] (l0) {$\lambda^{0}$};
\node[lPrimal,above=\stackgap of l0] (p0) {$\mathbf{p}^{0}$};
\node[lDual,below=\stackgap of l0]   (n0) {$\bm{\nu}^{0}$};
\coordinate (b0x) at ($(l0.east)+(0.5pt,0)$);
\draw[brc] ($(b0x |- p0.north)$) -- ($(b0x |- n0.south)$)
  node[midway,right=2.6pt,zlab] (z0) {$\mathbf{z}^{0}$};

\node[Tblk,right=of z0] (T1) {$\mathcal{T}$};
\node[zlab,right=of T1] (z1) {$\mathbf{z}^{1}$};
\node[Tblk,right=of z1] (T2) {$\mathcal{T}$};
\node[zlab,right=of T2] (z2) {$\mathbf{z}^{2}$};
\node[zlab,right=of z2] (dots) {$\cdots$};
\node[Tblk,right=of dots] (TK) {$\mathcal{T}$};
\node[zlab,right=of TK] (zs) {$\hat{\mathbf{z}}$};

\coordinate (bsx) at ($(zs.east)+(1.2pt,0)$);
\node[lDual,anchor=west] (ls) at ($(bsx)+(3.1pt,0)$) {$\hat{\lambda}$};
\node[lPrimal,above=\stackgap of ls] (ps) {$\hat{\mathbf{p}}$};
\node[lDual,below=\stackgap of ls]   (ns) {$\hat{\bm{\nu}}$};
\draw[brc] ($(bsx |- ns.south)$) -- ($(bsx |- ps.north)$);

\foreach \a/\b in {z0/T1, T1/z1, z1/T2, T2/z2, z2/dots, dots/TK, TK/zs}
  \draw[wState,->] (\a) -- (\b);

\ifshowdata
  \coordinate (bus) at ($(T1.south)+(0,-\busdrop)$);
  \node[lConst,anchor=east] (dc) at ($(bus -| z0.west)+(0,0)$) {$(\mathbf{d},c)$};
  \draw[wConst] (dc.east) -- (bus -| TK);
  \foreach \t in {T1,T2,TK} \draw[wConst,->] (bus -| \t) -- (\t.south);
\fi
\ifshowtheta
  \coordinate (tbus) at ($(T1.north)+(0,\thetarise)$);
  \node[lLearn,anchor=east] (th) at (tbus -| z0.west) {$\bm{\theta}$};
  \draw[wLearn] (th.east) -- (tbus -| TK);
  \foreach \t in {T1,T2,TK} \draw[wLearn,->] (tbus -| \t) -- (\t.north);
\fi
\end{tikzpicture}%
}%
        \subfloat[\label{fig:deq_chain}]{\resizebox{\columnwidth}{!}{\usebox0}}\par
        \vfill
        \sbox0{\edef\deqRestoreCatcodes{\catcode32=\the\catcode32\relax\catcode13=\the\catcode13\relax}%
\catcode32=9\relax\catcode13=9\relax
\definecolor{cPrimal}{HTML}{A64209}    %
\definecolor{cDual}{HTML}{1F6FB2}      %
\definecolor{cInk}{HTML}{253341}       %
\definecolor{cGray}{HTML}{52606D}      %
\definecolor{cLayerBg}{HTML}{DDEBD2}   %
\colorlet{cDualText}{cDual!80!black}
\definecolor{cAccent}{HTML}{7A5195}    %
\colorlet{cAccentText}{cAccent!92!black}
\def\fT{\fontsize{12}{14}\selectfont}    %
\def\fbig{\fontsize{10}{12}\selectfont}  %
\def\fvar{\fontsize{9}{11}\selectfont}   %
\def\fnote{\fontsize{8}{10}\selectfont}  %
\def\Tw{8mm}\def\Th{15mm}              %
\def\Sw{8mm}\def\Sh{7.5mm}             %
\def\initwire{10.5mm}                   %
\def\readwire{5mm}                     %
\def\kpwire{12mm}                      %
\def\exitwire{14mm}                    %
\def\looprise{4.6mm}                   %
\def\stackgap{3.8mm}                   %
\def\busdrop{5mm}                      %
\newif\ifshowdata \showdatatrue        %
\def\feedsep{1.8mm}                    %
\newif\ifshowtheta \showthetafalse     %
\deqRestoreCatcodes%
\begin{tikzpicture}[text=cInk,
  >={Stealth[length=2mm,width=1.6mm]},
  Tblk/.style={draw=cInk,line width=1.1pt,fill=cLayerBg,rounded corners=1pt,
               minimum width=\Tw,minimum height=\Th,inner sep=0pt,font=\fT},
  Sblk/.style={draw=cInk,line width=0.8pt,fill=white,rounded corners=1pt,
               minimum width=\Sw,minimum height=\Sh,inner sep=0pt,font=\fbig},
  zlab/.style={font=\fbig,text=cInk,inner sep=1.6pt},
  lPrimal/.style={font=\fvar,text=cPrimal,inner sep=0.6pt},
  lDual/.style={font=\fvar,text=cDualText,inner sep=0.6pt},
  lConst/.style={font=\fnote,text=cInk,inner sep=1pt},
  lNote/.style={font=\fnote,text=cInk,inner sep=1.2pt},
  wState/.style={cInk,line width=1.4pt},
  wInit/.style={cInk,line width=1.1pt,dash pattern=on 2.6pt off 1.8pt},
  wConst/.style={cGray,line width=0.95pt},
  wLearn/.style={cAccent,line width=0.95pt},
  lLearn/.style={font=\fnote,text=cAccentText,inner sep=1pt},
  dot/.style={circle,fill=cInk,inner sep=0pt,minimum size=4pt},
  brc/.style={decorate,decoration={brace,amplitude=2.6pt},cInk,line width=0.75pt}]

\node[lDual] (l0) {$\lambda^{0}$};
\node[lPrimal,above=\stackgap of l0] (p0) {$\mathbf{p}^{0}$};
\node[lDual,below=\stackgap of l0]   (n0) {$\bm{\nu}^{0}$};
\coordinate (b0x) at ($(l0.east)+(0.5pt,0)$);
\draw[brc] ($(b0x |- p0.north)$) -- ($(b0x |- n0.south)$)
  node[midway,right=2.6pt,zlab] (z0) {$\mathbf{z}^{0}$};

\node[Sblk,anchor=west] (S) at ($(z0.east)+(\initwire,0)$) {$\mathbf{z}^{k}$};
\node[Tblk,anchor=west] (T) at ($(S.east)+(\readwire,0)$) {$\mathcal{T}$};
\coordinate (jout) at ($(T.east)+(\kpwire,0)$);                  %
\node[zlab,anchor=west] (zs) at ($(jout)+(\exitwire,0)$) {$\hat{\mathbf{z}}$};

\coordinate (bsx) at ($(zs.east)+(1.2pt,0)$);
\node[lDual,anchor=west] (ls) at ($(bsx)+(3.1pt,0)$) {$\hat{\lambda}$};
\node[lPrimal,above=\stackgap of ls] (ps) {$\hat{\mathbf{p}}$};
\node[lDual,below=\stackgap of ls]   (ns) {$\hat{\bm{\nu}}$};
\draw[brc] ($(bsx |- ns.south)$) -- ($(bsx |- ps.north)$);

\draw[wInit,->] (z0) -- node[above=2pt,lNote] {$k=0$} (S);
\draw[wState,->] (S) -- (T);
\draw[wState] (T) -- node[above=0.5pt,zlab] {$\mathbf{z}^{k+1}$} (jout);
\coordinate (ltop) at ($(T.north)+(0,\looprise)$);
\draw[wState,->,rounded corners=2.5pt] (jout) |- (ltop) -| (S.north);
\node[lNote,above=0.5pt] at (ltop) {$k\leftarrow k+1$};
\draw[wState,->] (jout) -- node[below=1pt,lNote] {converged} (zs);
\node[dot] at (jout) {};

\coordinate (bus) at ($(T.south)+(0,-\busdrop)$);
\ifshowtheta \coordinate (inDC) at ($(T.south)+(-\feedsep,0)$);
\else        \coordinate (inDC) at (T.south); \fi
\ifshowdata
  \node[lConst,anchor=east] (dc) at ($(bus -| z0.west)$) {$(\mathbf{d},c)$};
  \draw[wConst] (dc.east) -- (bus -| inDC);
  \draw[wConst,->] (bus -| inDC) -- (inDC);
\fi
\ifshowtheta
  \coordinate (inTh) at ($(T.south)+(\feedsep,0)$);
  \node[lLearn,anchor=west] (th) at (bus -| zs.west) {$\bm{\theta}$};
  \draw[wLearn] (th.west) -- (bus -| inTh);
  \draw[wLearn,->] (bus -| inTh) -- (inTh);
\fi
\end{tikzpicture}%
}%
        \subfloat[\label{fig:deq_loop}]{\resizebox{\columnwidth}{!}{\usebox0}}}}%
    \sbox{\figleftbox}{\raisebox{\dp\figleftbox}{\usebox{\figleftbox}}}%
    \usebox{\figleftbox}\hfill\usebox{\figcolbox}%
    \setcounter{subfigure}{3}\setcounter{subfigure@save}{3}%
    \fi
    \caption{(a)~Loop-OPF as a stack of identical layers $\Tmap$; (b)~its equivalent looped form; (c)~one layer $\Tmap$. Topology-dependent quantities carry the superscript $(c)$, and the step sizes are in purple.}
    \label{fig:solver_layer}
\end{figure*}

We develop Loop-OPF, a \textit{solver-informed} learning-for-OPF model that computes OPF solutions with \textit{guaranteed optimality} and  \textit{inference-time adaptivity} to unseen topologies. The core idea of Loop-OPF is to use a neural model to mimic the recursive updates in~\eqref{eq:pdhg} that can obtain the optimal solution, unlike the feedforward prediction approach adopted by the learning-for-OPF literature{\mohamad~\cite{kotary2021survey, fioretto2020acopf, singh2021sidnn, pan2019deepopf, donti2021dc3, chen2023e2elr, huang2024deepopfngt}}.
Once trained, our Loop-OPF model can be immediately used for any  {\mohamad new scenario $c$, such as a line or generator outage,} and at any inference cost. We first present the architecture of Loop-OPF, followed by discussing its optimality guarantees and topology adaptivity. 

Fig.~\ref{fig:solver_layer} illustrates the overall architecture of Loop-OPF. As shown in  Fig.~\ref{fig:deq_chain}, the neural model consists of multiple, \textit{identical} $\Tmap$ layers that map the initial primal-dual variables stacked in $\bbz^{0}=(\bbp^{0},\lambda^{0},\bbnu^{0})$ to the final prediction $\hat{\bbz}$. Each layer $\Tmap$, as detailed in Fig.~\ref{fig:pdhg_layer}, performs one full PDHG iteration~\eqref{eq:pdhg}. Specifically,  it incorporates two given input parameters used by \eqref{eq:dcopf}: the demand $\bbd$  and {updated parameters for a new OPF scenario $c$. For the latter, any OPF scenario $c$ can update $\bbH^{(c)}$ and $\bbg^{(c)}(\bbd)$ of \eqref{eq:doubled_flow} as in the post-line-outage constraints \eqref{eq:scopf_cont},  or change $[\underline{\bbp}^{(c)},\bar{\bbp}^{(c)}]$ to match the outaged generator.}

 Hence, the Loop-OPF layer $\Tmap$ directly incorporates any new OPF instance to facilitate the topology adaptivity, as detailed later. For the $k$-th layer, it can be expressed as
\begin{equation}
\bbz^{k+1} = \Tmap(\bbz^{k};\bbd,c).
\label{eq:deq_operator}
\end{equation}
  Each layer $\Tmap$ is parameterized by the step-sizes in \eqref{eq:pdhg}: $\sigma_{\lambda}$, $\bbD_{\bbsigma}$ and
$\bbD_{\bbtau}$, as shown in purple in Fig.~\ref{fig:pdhg_layer}.  They constitute as the only learnable weight parameters for Loop-OPF. In Section~\ref{sec:training}, the training procedure stacks these step-sizes into a single vector $\bbtheta$ of learnable parameters and writes $\Tmap_{\bbtheta}$ to denote this dependence.
At the equilibrium of \eqref{eq:deq_operator}, the final prediction satisfies the \textit{fixed-point} condition  for \eqref{eq:pdhg}:
\begin{equation}
\hat{\bbz} = \Tmap(\hat{\bbz};\bbd,c).
\label{eq:deq_equilibrium}
\end{equation}
Throughout, $\hat{\bbz}=(\hat{\bbp},\hat{\lambda},\hat{\bbnu})$ denotes the Loop-OPF output, while $\bbz^{\star}=(\bbp^{\star},\lambda^{\star},\bbnu^{\star})$ is reserved for the optimal solution of~\eqref{eq:dcopf} and its multipliers.

Our neural model can be equivalently viewed as a loop transformation, thus the name Loop-OPF. Hence, Fig.~\ref{fig:deq_loop} shows a single layer $\Tmap$ is applied repeatedly in a feedback form. Indeed, it is sufficient to use this compact, single-layer structure to fully characterize the overall neural model, instead of the multi-layer representation in Fig.~\ref{fig:deq_chain}. This compact structure mirrors the parameter-efficient and scalable loop neural models used in language modeling~\cite{yang2024looped}, enabling  efficient model training as discussed in Section \ref{sec:training}.

\subsection{Guaranteed Optimality/Feasibility}
\label{sec:optimality}

The key advantage of our Loop-OPF design lies in the \textit{guaranteed optimality} of  $\hat{\bbz}$. Under the standard conditions on convex DC-OPF collected in Assumption~\ref{as:convex}, we formally state this result in Proposition~\ref{prop:kkt}. %
 \begin{assumption}[Convexity and regularity]
    \label{as:convex}
The problem~\eqref{eq:dcopf} satisfies the following conditions: (i) \textit{convex} $f$ with $\bbQ \succeq 0$; (ii) \textit{feasible} constraint set with $\bbd$ in the operating range; (iii) \textit{compact} range $[\underline{\bbp},\bar{\bbp}]$; and (iv) Slater's condition, i.e., there exists a feasible $\tilde{\bbp}$ in the interior of $[\underline{\bbp},\bar{\bbp}]$ attaining the strict inequality $\bbH\tilde{\bbp} < \bbg(\bbd)$.
\end{assumption}
    
\begin{proposition}[Optimality of $\hat{\bbz}$]
    \label{prop:kkt}
Under Assumption~\ref{as:convex}, if the sequence $\{\bbz^{k}\}$ generated by~\eqref{eq:deq_operator} converges to its fixed point
$\hat{\bbz}$ in \eqref{eq:deq_equilibrium}, then 
    the dispatch $\hat{\bbp}=\bbp^{\star}$ is the  global
    minimizer of~\eqref{eq:dcopf}, with $(\hat{\lambda},\hat{\bbnu})=(\lambda^{\star},\bbnu^{\star})$ for
    the corresponding multipliers.
    \end{proposition}
    
Intuitively, the equilibrium condition in~\eqref{eq:pdhg} is exactly a KKT point of~\eqref{eq:dcopf}, for which convexity and Slater's condition in Assumption~\ref{as:convex} jointly ensure the global optimality~\cite{boyd2004convex}.  Note that this  guarantee holds for any convex $f$, including a piecewise linear cost~\cite{chambolle2011pdhg, coffrin2021pwlcost}. This result makes optimality a property of the Loop-OPF architecture itself, regardless of the training. Consequently, Loop-OPF is guaranteed to attain optimality at convergence, and learning only affects how fast $\hhatbbz$ is reached. This also simplifies training: since $\hhatbbz$ already satisfies all OPF constraints, it is not necessary to use the training to enforce them, as discussed in Section~\ref{sec:training}.

\subsection{Inference-Time Topology Adaptivity to Unseen Scenarios}
\label{sec:topology}

As the layer $\Tmap$ mimics the PDHG updates, the Loop-OPF model, once trained, can be updated with any new topology or OPF scenario $c$ at inference.  
{\mohamad Such a scenario $c$ can be a line outage, which changes $\bbH^{(c)}$ and $\bbg^{(c)}(\bbd)$, or a generator outage, or any other change in the generator limits, which only modifies $[\underline{\bbp}^{(c)},\bar{\bbp}^{(c)}]$.} %
As Fig.~{\mohamad\ref{fig:pdhg_layer}} shows, %
 {\mohamad the demand $\bbd$ and these scenario data enter $\Tmap$ directly at every pass of the loop.}  {\mohamad Since none of these data affect the learnable parameters, a trained Loop-OPF model is updated with a new scenario $c$ by changing only its input data, while its learned parameters stay unchanged.} %
 This property essentially enables the inference-time adaptivity of Loop-OPF \textit{without any retraining}, unlike learning-for-OPF models  {\mohamad whose weights absorb} the topology seen in training. %
{\mohamad With this adaptivity, Proposition~\ref{prop:kkt} readily applies to the OPF problem of any new scenario, as stated below.
\begin{corollary}[Optimality of $\hat{\bbz}$ for new scenario]
    \label{cor:new_scenario}
If Assumption~\ref{as:convex} holds for the new OPF problem under scenario $c$, the fixed point of the updated Loop-OPF model $\Tmap(\cdot\,;\bbd,c)$ is still an optimal solution of that problem, regardless of the model parameters.
\end{corollary}} %

{\mohamad This adaptivity also allows Loop-OPF to learn its parameters from samples drawn from \textit{multiple scenarios}, rather than from a single topology. Each training sample pairs a demand $\bbd$ with a scenario $c$, and the PDHG iterations solve the corresponding OPF instance. A single set of parameters $\bbtheta$ is then learned to minimize the expected loss over all sampled scenarios. This procedure can be viewed from a \textit{meta-learning} perspective~\cite{finn2017maml}: each scenario defines a task, the PDHG iterations form an inner loop that solves it, and training forms an outer loop that learns $\bbtheta$ across tasks.  {\mohamad Related meta-learning methods also learn the step sizes or update rule of an inner optimizer, but that optimizer is typically gradient descent for training a neural network~\cite{andrychowicz2016learning}. In Loop-OPF, the inner optimizer is the PDHG solver of the OPF problem itself, so by Corollary~\ref{cor:new_scenario} the learned $\bbtheta$ changes only how fast Loop-OPF converges, not the optimal solution it converges to.} %
 Hence, training over many scenarios encourages $\bbtheta$ to speed up convergence for a wide range of scenarios, including unseen ones. Section~\ref{sec:results_topology_conditioned} confirms this on test outages that do not appear in training.} %

Although the optimality of $\hat{\bbz}$ is established for Loop-OPF, Proposition \ref{prop:kkt} does not provide the convergence conditions or control the \textit{convergence rate}. %
To ensure well-conditioned  convergence, we put forth a learning framework to optimize the step-sizes from data scenarios. Due to fast-varying grid operations, it is critical to choose appropriate step-sizes to accelerate the convergence rate while not overshooting it \cite{goldstein2015adaptive, applegate2021pdlp}. One important aspect is finding step-sizes that can adapt to a wide range of operating conditions under generator/line outages, commonly in daily grid operations. To achieve this goal, our Loop-OPF learns the set of best step-sizes over multiple topology scenarios, such that no retraining is needed  {\mohamad for a new scenario}. Furthermore, our Loop-OPF, once trained,  can be
deployed under any available latency budget in the inference-stage, as discussed below.

\vspace{3pt}
\begin{remark}[Anytime-use Inference]
\label{rem:anytime}
A key characteristic of our solver-informed Loop-OPF is that it can be used under any level of available inference budget, making it an \textit{anytime}
method~\cite{zilberstein1996anytime}.  For a given time budget, it is possible to perform the iterations \eqref{eq:deq_operator} as many as the budget allows. Note that this flexibility does not apply to common
learning-for-OPF methods, which have a fixed inference-stage procedure and cannot adapt to any time budget.
Of course, the accuracy of Loop-OPF inference improves as the budget grows, as detailed in Section~\ref{sec:results_fixedk}. 
\end{remark}

\section{Memory-Efficient Training of Loop-OPF}
\label{sec:training}

To optimize the step sizes, our training uses the \textit{self-supervised learning (SSL)} approach for computational efficiency. Unlike supervised learning, the SSL framework does not require generating thousands of target labels a
priori~\cite{park2023pdl, park2024pdlscopf, chen2023e2elr}, eliminating the offline cost of solving the OPF instance for every training sample. It also makes it inexpensive to train on samples from multiple contingency scenarios.  {\mohamad To add a scenario $c$ to the training set, we only need its data $\bbH^{(c)}$, $\bbg^{(c)}(\bbd)$ and $[\underline{\bbp}^{(c)},\bar{\bbp}^{(c)}]$, which enter $\Tmap$ directly (Section~\ref{sec:topology}); no OPF has to be solved.}

Without any labels, we use the following loss function as a weighted sum of the OPF cost and constraint violations defined for any Loop-OPF prediction $\hat{\bbp}(\bbd)$:
\begin{align}
\Jssl(\hat{\bbp}; \bbd)
&= \frac{w_{\mathrm{cost}}}{\mathbf{1}^\top\bbd}\, f(\hat{\bbp})
 + w_{\mathrm{bal}}\bigl(\mathbf{1}^\top\hat{\bbp} - \mathbf{1}^\top\bbd\bigr)^{2} \nonumber\\
&\quad + w_{\mathrm{flow}}\bigl\|\bigl(\bbH\hat{\bbp} - \bbg(\bbd) - \varepsilon_{\mathrm{flow}}\mathbf{1}\bigr)_+\bigr\|_{2}^{2},
\label{eq:selfsup_main}
\end{align}
where the hyperparameters $w_{\mathrm{cost}}$, $w_{\mathrm{bal}}$ and
$w_{\mathrm{flow}}$ balance the cost against the other two penalty terms. The cost is divided by the total demand $\mathbf{1}^\top\bbd$, so that the cost term has a comparable scale across load levels, and $\varepsilon_{\mathrm{flow}}\geq 0$ is a small tolerance: only flow-limit violations larger than $\varepsilon_{\mathrm{flow}}$ are penalized. 
One key difference between Loop-OPF and the existing SSL-based OPF approaches~\cite{donti2021dc3, park2023pdl} is that our Loop-OPF \textit{does not rely solely on} the two penalty terms  in \eqref{eq:selfsup_main} to enforce the OPF constraints. Recall that Proposition~\ref{prop:kkt} already guarantees that the converged $\hhatbbz$ is feasible and optimal, and thus both penalty terms vanish at the prediction output $\hhatbbp$. Although the penalty terms are not necessary to drive the output $\hhatbbp$ toward
constraint satisfaction, they provide better numerical performance for the training process. 
 {\mohamad In practice, training runs a finite number of iterations, so the output satisfies the constraints only approximately. The penalty terms then push the learned step sizes toward outputs that violate the constraints less, and after fewer iterations.}

\vspace{3pt}
\noindent \textbf{Forward Pass.} 
\label{sec:reparam}
The simplest way to execute the forward pass follows
Fig.~\ref{fig:deq_chain}: starting from an initial estimate $\bbz^{0}$, one can apply $\Tmap$
repeatedly as in~\eqref{eq:deq_operator} until the iterates reach the fixed-point $\hhatbbz$
of~\eqref{eq:deq_equilibrium}. This boils down to an iterative
optimization method, and the resultant multi-layer, recursive model is known as an \textit{unrolled neural network}~\cite{li2024pdhgnet}. Building on the loop view introduced in Section~\ref{sec:method}, a more numerically efficient forward pass treats $\hhatbbz$ directly as the fixed point of $\Tmap$ and computes it with a fixed-point solver, without rolling out the recursive layers. We use a Picard warm-start followed by
the Anderson acceleration~\cite{walker2011anderson} for this fixed-point computation, with more details in
Section~\ref{sec:experiments}.

\vspace{3pt}
\noindent \textbf{Backward Pass.} The backward pass computes the gradient of the loss~\eqref{eq:selfsup_main}
with respect to the step-sizes. One key strategy that we have adopted is a transform step for the step-sizes for an enhanced convergence of Loop-OPF. Specifically, we parameterize the step-sizes so that, after every backward update, they still satisfy the sufficient condition for
the convergence of PDHG iterations~\cite{chambolle2011pdhg, pock2011precond}. Specifically, by defining $\bar{\bbD}_{\bbsigma}:=\diag(\sigma_{\lambda},\bbsigma)$, the sufficient condition requires
\begin{equation}
\bigl\|\bar{\bbD}_{\bbsigma}^{1/2}\,\bbK\,\bbD_{\bbtau}^{1/2}\bigr\| < 1
\label{eq:admissible}
\end{equation}
where the matrix $\bbK = [\mathbf{1},\ \bbH^{\top}]^{\top}\in\R^{(1+2L)\times N}$ here stacks the
matrices of~\eqref{eq:dcopf_bal}-\eqref{eq:dcopf_flow} while 
$\|\cdot\|$ denotes the spectral norm. To maintain \eqref{eq:admissible} throughout the backward pass, we introduce the new parameter vector $\bbtheta=(\rho,s,\bba,\bbb)$ for each Loop-OPF layer $\Tmap$,  which can be used to form the original step-sizes as 
\begin{equation}
\bbD_{\bbtau} = \frac{\sqrt{\rho}}{\kappa}\,e^{s}\diag\!\bigl(e^{\bba}\bigr),
\qquad
\bar{\bbD}_{\bbsigma} = \frac{\sqrt{\rho}}{\kappa}\,e^{-s}\diag\!\bigl(e^{\bbb}\bigr),
\label{eq:step_param_main}
\end{equation}
 with the coefficient $\kappa := \bigl\|\diag(e^{\bbb/2})\,\bbK\,\diag(e^{\bba/2})\bigr\|$.
This change of variables satisfies~\eqref{eq:admissible} for any $0<\rho<1$,
since the left-hand side of~\eqref{eq:admissible} equals to $\rho$ for every
$\bbtheta$. This way, the backward pass no longer needs any expensive projection to satisfy \eqref{eq:admissible}.  Moving forward, we use $\Tmap_{\bbtheta}$ for the operator $\Tmap$
of~\eqref{eq:deq_operator} with learnable parameters in $\bbtheta$.

To obtain the loss gradient, the backward pass leverages the \textit{implicit differentiation} directly on the equilibrium
condition~\eqref{eq:deq_equilibrium}. This approach eliminates the need of  backpropagating
through multiple iterations of~\eqref{eq:deq_operator}, required by the training of unrolled neural networks. This
approach follows from the \textit{implicit function theorem}~\cite{krantz2002ift}, as advocated by the deep equilibrium (DEQ) model training in~\cite{bai2019deq}. More concretely, both sides of
$\hat{\bbz} = \Tmap_{\bbtheta}(\hat{\bbz};\bbd,c)$ depend on $\bbtheta$, either directly through the parameterization or 
via the dependence of $\hat{\bbz}$.  Differentiating~\eqref{eq:deq_equilibrium} with respect to $\bbtheta$ gives rise to:
\begin{equation}
\frac{\partial \hat{\bbz}}{\partial \bbtheta}
= \frac{\partial \Tmap_{\bbtheta}}{\partial \bbz}\bigg|_{\hat{\bbz}}
  \frac{\partial \hat{\bbz}}{\partial \bbtheta}
+ \frac{\partial \Tmap_{\bbtheta}}{\partial \bbtheta}\bigg|_{\hat{\bbz}},
\label{eq:implicit_diff_raw}
\end{equation}
which leads to the following loss gradient:
\begin{equation}
\frac{\partial \hat{\bbz}}{\partial \bbtheta}
= \left(\bbI - \frac{\partial \Tmap_{\bbtheta}}{\partial \bbz}\bigg|_{\hat{\bbz}}\right)^{\!-1}
  \frac{\partial \Tmap_{\bbtheta}}{\partial \bbtheta}\bigg|_{\hat{\bbz}}.
\label{eq:implicit_diff_main}
\end{equation}
Clearly, the gradient computation here only need the value of fixed-point  $\hhatbbz$, at no need of any intermediate iterate $\bbz^{k}$. 

\vspace{3pt}
\begin{remark}[Memory-efficient Training]
\label{rem:adjoint}
Since the loss gradient~\eqref{eq:implicit_diff_main} requires only to know 
$\hhatbbz$, both the forward and backward passes benefit from efficient memory needs throughout the training process. The forward pass does not have to record the recursive updates of \eqref{eq:deq_operator} and can compute
$\hhatbbz$ with any fixed-point solver. Consequently, training memory is independent of the number of forward
iterations that are needed for convergence. Of course, the matrix $(\bbI-\partial \Tmap_{\bbtheta}/\partial\bbz)$ needs to be
nonsingular at $\hhatbbz$~\cite{krantz2002ift}, but this is typically easy to satisfy by using numerically stable solvers for the linear system in \eqref{eq:implicit_diff_main}. Unlike the unrolled network
in Fig.~\ref{fig:deq_chain} that has to store all iterates, the efficient training makes the proposed loop-OPF structure in Fig.~\ref{fig:deq_loop} very attractive for solving large-scale OPF problems. 
\end{remark}

\section{Numerical Tests}
\label{sec:results}

We evaluate Loop-OPF on four power networks with different levels of congestion. The evaluation has five studies. First, Section~\ref{sec:results_clean} compares the dispatch accuracy and constraint violations of Loop-OPF with those of learning-for-OPF models. It uses test samples drawn from the training scenario, without outages. Second, Section~\ref{sec:results_topology_conditioned} evaluates single line or generator outages. There, the learning-for-OPF models also receive the post-outage topology as input. Third, Section~\ref{sec:results_scopf} considers the preventive SC-OPF in Remark~\ref{rem:general}. In this problem, one dispatch must satisfy the network without outages and a set of line outages simultaneously. Fourth, Section~\ref{sec:results_solver} isolates the effect of training. It compares the trained and initial Loop-OPF models at the same iteration count. Last, Section~\ref{sec:results_fixedk} varies the number $K$ of iterations of the trained model. It validates that Loop-OPF remains accurate under a limited inference budget.

\subsection{Experimental Setup}
\label{sec:experiments}

\noindent  \textbf{Networks and data.} We use the IEEE 300-bus network~\cite{christie1993uwarchive} and the PEGASE 1354-, 2869-, and 9241-bus networks~\cite{fliscounakis2013pegase, josz2016pegase}. All four are distributed with MATPOWER~\cite{zimmerman2011matpower}. They are also included in the PGLib-OPF benchmark library~\cite{babaeinejad2019pglib}. Each network has its own dataset of $10{,}000$ samples. Of these, $7{,}000$ are used for training, $1{,}000$ for validation, and $2{,}000$ for testing. Each sample has either no outage, one line outage, or one generator outage. The outages are drawn from $32$ candidate lines and $32$ candidate generators in each network. {\mohamad The reference solution of each sample is computed with the DC-OPF solver of pandapower~\cite{thurner2018pandapower}, which uses a primal--dual interior-point method~\cite{wang2007mips}.} Each test set has about $500$ samples without outages, $740$ with a line outage, and $770$ with a generator outage. Table~\ref{tab:dataset} summarizes the networks.

\noindent  \textbf{Model and compute.} We train one Loop-OPF model per network. Each model learns a single pair $(\rho,s)$, shared by all samples of that network. This is the scalar case $\bba=\bbb=\mathbf{0}$ of~\eqref{eq:step_param_main}. Training runs for $30$ epochs {\mohamad with the Adam optimizer~\cite{kingma2015adam} in PyTorch~\cite{paszke2019pytorch}}. The learning rate is $10^{-4}$, and the flow penalty weight is $w_{\mathrm{flow}}=10^{8}$. Each forward pass runs at most $600$ iterations. We compute the fixed point $\hat{\bbz}$ of~\eqref{eq:deq_equilibrium} in two stages. A $50$-step Picard warm start~\cite{berinde2007iterative} is followed by Anderson acceleration with memory $m=4$~\cite{anderson1965iterative, walker2011anderson}. We report the checkpoint with the best validation performance. Thus, the test set plays no role in model selection. Training runs on NVIDIA GH200 nodes of the Vista supercomputer at the Texas Advanced Computing Center~\cite{tacc2026vista} and takes $8.7$ to $40.3$~h per network. Inference latency is measured on the same nodes, as the median run time of the trained model over repeated runs on the same input. {\mohamad The code is available at \url{https://github.com/MFHChehade/loop-opf}.}

\begin{table}[t]
\centering
\caption{Benchmark power networks.}
\label{tab:dataset}
\small
\setlength{\tabcolsep}{4pt}
\begin{tabular*}{\columnwidth}{@{}l@{\extracolsep{\fill}}c c c@{}}
\toprule
Network & Buses & Generators & Branches \\
\midrule
IEEE 300 & 300 & 69 & 411 \\
PEGASE 1354 & 1354 & 260 & 1991 \\
PEGASE 2869 & 2869 & 510 & 4582 \\
PEGASE 9241 & 9241 & 1445 & 16049 \\
\bottomrule
\end{tabular*}

\vspace*{-6pt}
\end{table}

\begin{table*}[t]
\centering
\caption{Learning-for-OPF models on test samples without outages. %
{\mohamad All models trained on samples without outages only.}}
\label{tab:learning_opf}
\small
\setlength{\tabcolsep}{2pt}
\begin{tabular*}{\textwidth}{@{}l@{\extracolsep{\fill}}rrrrrrrrrrrr@{}}
\toprule
 & \multicolumn{3}{c}{IEEE 300} & \multicolumn{3}{c}{PEGASE 1354} & \multicolumn{3}{c}{PEGASE 2869} & \multicolumn{3}{c}{PEGASE 9241} \\
\cmidrule(l{4pt}){2-4}\cmidrule(l{4pt}){5-7}\cmidrule(l{4pt}){8-10}\cmidrule(l{4pt}){11-13}
Method & {\mohamad RMSE} & {\mohamad Flow} & Gap & {\mohamad RMSE} & {\mohamad Flow} & Gap & {\mohamad RMSE} & {\mohamad Flow} & Gap & {\mohamad RMSE} & {\mohamad Flow} & Gap \\
\midrule
DeepOPF~\cite{pan2019deepopf} & {\mohamad 146.353} & {\mohamad 85.776} & 194.925 & {\mohamad $2.15{\times}10^{4}$} & {\mohamad 6{,}103.046} & $4.11{\times}10^{6}$ & {\mohamad $8.70{\times}10^{4}$} & {\mohamad 7{,}002.225} & $6.95{\times}10^{7}$ & {\mohamad $6.37{\times}10^{5}$} & {\mohamad $1.89{\times}10^{5}$} & $3.36{\times}10^{9}$ \\
DC3~\cite{donti2021dc3} & {\mohamad 143.443} & {\mohamad 0.328} & 197.474 & {\mohamad 515.968} & {\mohamad 4.428} & 2{,}369.618 & {\mohamad 1{,}531.471} & {\mohamad 9.504} & $2.14{\times}10^{4}$ & {\mohamad 8{,}327.106} & {\mohamad 2{,}705.967} & $5.73{\times}10^{5}$ \\
DeepLDE~\cite{kim2024deeplde} & {\mohamad 9.913} & {\mohamad 3.707} & 1.213 & {\mohamad 56.776} & {\mohamad \textbf{0.000}} & 32.429 & {\mohamad 182.833} & {\mohamad 4.589} & 314.448 & {\mohamad 389.996} & {\mohamad 9.844} & 1{,}336.807 \\
PDL~\cite{park2023pdl} & {\mohamad 0.845} & {\mohamad 0.624} & 0.119 & {\mohamad 0.248} & {\mohamad 0.246} & 0.012 & {\mohamad 0.020} & {\mohamad \textbf{0.000}} & \textbf{\textless 0.001} & {\mohamad 1.254} & {\mohamad 0.935} & 0.071 \\
GNN~\cite{owerko2020gnnopf} & {\mohamad 2.465} & {\mohamad 4.932} & 0.498 & {\mohamad 0.834} & {\mohamad 2.024} & 0.082 & {\mohamad 0.020} & {\mohamad \textbf{0.000}} & \textbf{\textless 0.001} & {\mohamad 0.192} & {\mohamad 2.304} & 0.069 \\
\addlinespace[2pt]
Loop-OPF & {\mohamad \textbf{\textless 0.001}} & {\mohamad \textbf{0.001}} & \textbf{\textless 0.001} & {\mohamad \textbf{\textless 0.001}} & {\mohamad \textless 0.01} & \textbf{\textless 0.001} & {\mohamad \textbf{\textless 0.001}} & {\mohamad \textbf{0.000}} & \textbf{\textless 0.001} & {\mohamad \textbf{0.051}} & {\mohamad \textbf{0.319}} & \textbf{0.017} \\
\bottomrule
\end{tabular*}

\end{table*}

\noindent \textbf{Metrics.} Dispatch accuracy is the global root-mean-square error (RMSE) against the reference solution. {\mohamad We report it as a percentage of the average generator output. The average generator output is the total generation divided by the number of generators, averaged over the test samples.} The flow violation measures line-flow feasibility. {\mohamad For each test sample, we take the \emph{worst} branch overload as a percentage of that branch's rating, and then average over the test samples.} {\mohamad Both normalizations make the values comparable across networks of different sizes.} The balance error is the mean absolute mismatch between generation and demand, in MW. Finite-iteration outputs can retain a
small balance error. If exact balance is required, a post-processing step can
remove it without retraining~\cite{donti2021dc3, chen2023e2elr}.
The cost gap is the mean absolute relative gap to the optimal cost of the reference solution. In the tables, Flow, Bal., and Gap denote the flow violation, balance error, and cost gap. {\mohamad RMSE, Flow, and Gap are in \%, and Bal. is in MW. The figures report RMSE and flow violation in MW.} Lower is better throughout.
\subsection{Comparison with Learning-for-OPF Models}
\label{sec:results_clean}

We compare Loop-OPF with five state-of-the-art learning-for-OPF models.
\textbf{DeepOPF}~\cite{pan2019deepopf} directly predicts generator dispatch.
\textbf{DC3}~\cite{donti2021dc3} predicts part of the dispatch and solves the power balance for the rest. It then takes a few gradient steps to reduce the violations of generator and line-flow limits.
\textbf{DeepLDE}~\cite{kim2024deeplde} trains a dispatch predictor using constraint penalties whose dual weights are updated during training.
\textbf{PDL}~\cite{park2023pdl} instead trains separate networks to predict dispatch
and the associated constraint multipliers.
The \textbf{GNN} model~\cite{owerko2020gnnopf} combines features from connected buses to
predict generator outputs. {All six models {\mohamad (DeepOPF, DC3, DeepLDE, PDL, GNN, and Loop-OPF)} are trained and validated only on the samples without outages of each network. Each model uses a single seed and the checkpoint chosen by the validation score{\mohamad, except for the GNN, which keeps its final weights}. The baselines keep their original IEEE-300 budgets on every network. DeepOPF uses $200$ epochs, a batch size of $64$, SGD with momentum $0.9$, a learning rate of $10^{-4}$, and a $4h$-$2h$-$h$ architecture with $h=8$. DeepLDE and PDL use $1{,}000$ epochs, a batch size of $64$, and a learning rate of $10^{-4}$. DC3 uses its original $1{,}000$-epoch configuration. {\mohamad The GNN uses its original $200$ epochs, a learning rate of $10^{-3}$, and $3$ layers of $64$ hidden units.} DeepOPF and DC3 are DC-OPF adaptations of the authors' code. DeepLDE and PDL are adapted to DC-OPF from their original AC-OPF code.  {\mohamad Loop-OPF is trained for only $30$ epochs. All six models are evaluated on exactly the same set of test samples without outages.}} %
Finally, unlike many other methods, Loop-OPF uses no post-processing step to enforce power balance. If needed, the post-processing step of Section~\ref{sec:experiments} can restore its exact balance.

Table~\ref{tab:learning_opf} shows that Loop-OPF attains the lowest RMSE and the lowest (or tied lowest) cost gap on all four networks. It also has the lowest flow violation on IEEE 300 and PEGASE 9241. PDL is the strongest baseline. Its RMSE is  {\mohamad more than $700$} times that of Loop-OPF on PEGASE 1354 and  {\mohamad $25$} times on PEGASE 9241. {\mohamad On PEGASE 1354, Loop-OPF is trained for $12$ epochs with at most $2{,}400$ iterations per forward pass. Its RMSE and flow violation are $0.45$ and $4.09$~MW after $2{,}400$ iterations, and $0.001$ and $0.004$~MW after the $10{,}000$ used in Table~\ref{tab:learning_opf}.} %
 On this network, DeepLDE has no flow violation. However, its RMSE is {\mohamad $56.8\%$}, its cost gap is $32\%$, and its generator limits are violated by $79.7$~MW on average. DeepOPF and DC3 show the worst performance on every network. Their cost gaps are above $190\%$, and they violate the generator limits by large amounts. On PEGASE 2869, PDL, GNN, and Loop-OPF all have no flow violation. This network is \textit{largely uncongested}, so it is not a strong test case for congested lines.

\subsection{Comparison under Outage Scenarios}
\label{sec:results_topology_conditioned}

\begin{table}[t]
\centering
\caption{Topology-conditioned models on test samples with and without outages.}
\label{tab:topology_conditioned}
\small
\setlength{\tabcolsep}{1.5pt}
\begin{tabular*}{\columnwidth}{@{}ll@{\extracolsep{\fill}}rrrr@{}}
\toprule
Network & Method & {\mohamad RMSE} & {\mohamad Flow} & Bal. & Gap \\
 & & {\mohamad (\%)} & {\mohamad (\%)} & (MW) & (\%) \\
\midrule
\multirow{3}{*}{IEEE 300}
 & Adm.\ dense~\cite{falconer2023topology} & {\mohamad 5.383} & {\mohamad 9.939} & 0.004 & 0.692 \\
 & Dyn.\ GNN~\cite{liu2023topognn} & {\mohamad 6.245} & {\mohamad 6.207} & \textless 0.001 & 1.066 \\
 & Loop-OPF & {\mohamad\textbf{0.003}} & {\mohamad\textbf{\textless 0.001}} & \textbf{\textless 0.001} & \textbf{\textless 0.001} \\
\addlinespace[3pt]
\multirow{3}{*}{\shortstack[l]{PEGASE\\1354}}
 & Adm.\ dense~\cite{falconer2023topology} & {\mohamad 1.722} & {\mohamad 2.768} & 0.062 & 0.188 \\
 & Dyn.\ GNN~\cite{liu2023topognn} & {\mohamad 2.874} & {\mohamad 4.461} & \textbf{0.016} & 0.119 \\
 & Loop-OPF & {\mohamad\textbf{0.847}} & {\mohamad\textbf{1.601}} & 0.031 & \textbf{0.042} \\
\addlinespace[3pt]
\multirow{3}{*}{\shortstack[l]{PEGASE\\2869}}
 & Adm.\ dense~\cite{falconer2023topology} & {\mohamad 0.196} & {\mohamad 0.046} & 0.253 & \textless 0.001 \\
 & Dyn.\ GNN~\cite{liu2023topognn} & {\mohamad 0.212} & {\mohamad 0.047} & 0.253 & \textless 0.001 \\
 & Loop-OPF & {\mohamad\textbf{0.002}} & {\mohamad\textbf{0.000}} & \textbf{0.002} & \textbf{\textless 0.001} \\
\addlinespace[3pt]
\multirow{3}{*}{\shortstack[l]{PEGASE\\9241}}
 & Adm.\ dense~\cite{falconer2023topology} & {\mohamad 2.403} & {\mohamad 2.547} & 0.988 & 0.168 \\
 & Dyn.\ GNN~\cite{liu2023topognn} & {\mohamad 56.252} & {\mohamad 17.609} & 0.252 & 31.954 \\
 & Loop-OPF & {\mohamad\textbf{0.056}} & {\mohamad\textbf{0.354}} & \textbf{0.147} & \textbf{0.020} \\
\bottomrule
\end{tabular*}

\end{table}

To evaluate performance under outages, we further compare Loop-OPF with two topology-conditioned learning-for-OPF models. Both models receive the post-outage topology as input. The admittance-conditioned dense network~\cite{falconer2023topology} takes as input the demand, generator availability, active-line admittances, and line-outage response features. The dynamic GNN~\cite{liu2023topognn} uses active line connections to combine features from neighboring buses before predicting generator outputs. Both models predict the dispatch and then apply a bounded power-balance repair. Each model is trained separately on each network for $100$ epochs on the $7{,}000$ training samples. It is then tested on the same $2{,}000$ test samples as Loop-OPF. These include samples with and without outages. The outages in the test set do not appear in training. On each network, the $7$ line and $7$ generator outages of the test set are distinct from the $22$ line and $22$ generator outages used for training. The reference solution of each sample is the OPF solution for its own demand and its own post-outage network. This study thus differs from the preventive SC-OPF of Section~\ref{sec:results_scopf}. There, one dispatch must satisfy several outages at once.

Table~\ref{tab:topology_conditioned} shows that Loop-OPF has the lowest
RMSE, mean worst flow violation, and cost gap on all four networks. {\mohamad The flow violation of Loop-OPF is at most $1.6\%$ (PEGASE 1354). The topology-conditioned models reach $2.5\%$ to $17.6\%$ on the three congested networks. The RMSE of Loop-OPF is at most $0.85\%$, against $1.7\%$ to $56\%$ for these models on the same networks.} %
As in Section~\ref{sec:results_clean}, the largely uncongested PEGASE 2869 is not a strong test case for congested lines.

\subsection{Preventive Security-Constrained (SC-)OPF}
\label{sec:results_scopf}

\begin{figure}[t]
    \centering
    \def\scopfcsv{figures/fig_scopf/scopf.csv}%
    \definecolor{scRaw}{HTML}{F6CFCB}%
\definecolor{scFix}{HTML}{E9A6A1}%
\definecolor{scHatch}{HTML}{B64342}%
\definecolor{scLoop}{HTML}{0F4D92}%
\definecolor{scAxis}{HTML}{253341}%
\def\scopfW{31.5mm}%
\def\scopfH{45mm}%
\def\scopfBar{6.5pt}%
\ifdefined\scopfcsv\else\def\scopfcsv{figures/fig_scopf/scopf.csv}\fi
\pgfplotstableread[col sep=comma]{\scopfcsv}\scopftab%
\begin{tikzpicture}[font=\footnotesize, text=scAxis]
\pgfplotsset{
  sc bar/.style n args={3}{ybar, bar width=\scopfBar, bar shift=#1, fill=#2, draw=black, line width=0.45pt, #3},
  sc hatch/.style={postaction={pattern=north east lines, pattern color=scHatch}},
}
\def\scLab#1#2#3#4{\node[rotate=90, anchor=west, font=\scriptsize, text=scAxis, inner sep=1pt] at ([xshift=#3]axis cs:#1,#2) {#4};}
\begin{groupplot}[
    group style={group size=2 by 1, horizontal sep=13mm},
    scale only axis, width=\scopfW, height=\scopfH,
    xmin=0.45, xmax=3.55, xtick={1,2,3},
    xticklabels={{Deep-\\OPF},{Constr.-\\driven},{Loop-\\OPF}},
    xticklabel style={align=center, font=\footnotesize, text=scAxis},
    tick label style={font=\footnotesize, text=scAxis},
    label style={font=\footnotesize, text=scAxis},
    title style={font=\footnotesize, text=scAxis, inner sep=0pt, yshift=0.4mm},
    axis lines*=left,
    axis line style={scAxis, line width=0.6pt},
    tick style={scAxis, line width=0.6pt},
    major x tick style={draw=none},
    major tick length=2.5pt, minor tick length=0pt,
    unbounded coords=jump, enlarge x limits=false,
]
\nextgroupplot[ymin=0, ymax=128, ytick={0,50,100}, ylabel={secure samples (\%)}, title={(a)}]
  \addplot[sc bar={-4.6pt}{scRaw}{}] table[x=x, y=secure_asis, meta=lab_secure_asis] {\scopftab};
  \addplot[sc bar={4.6pt}{scFix}{sc hatch}] table[x=x, y=secure_repair, meta=lab_secure_repair] {\scopftab};
  \addplot[sc bar={0pt}{scLoop}{}] table[x=x, y=secure_loop, meta=lab_secure_loop] {\scopftab};
  \scLab{1}{1.40}{-4.6pt}{1.4}\scLab{1}{100}{4.6pt}{100}
  \scLab{2}{15.70}{-4.6pt}{15.7}\scLab{2}{100}{4.6pt}{100}
  \scLab{3}{100}{0pt}{100}
\nextgroupplot[ymode=log, ymin=1e-7, ymax=3000, ytick={1e-6,1e-4,1e-2,1}, log origin=infty, ylabel={cost gap (\%)}, title={(b)}]
  \addplot[sc bar={-4.6pt}{scRaw}{}] table[x=x, y=gap_asis, meta=lab_gap_asis] {\scopftab};
  \addplot[sc bar={4.6pt}{scFix}{sc hatch}] table[x=x, y=gap_repair, meta=lab_gap_repair] {\scopftab};
  \addplot[sc bar={0pt}{scLoop}{}] table[x=x, y=gap_loop, meta=lab_gap_loop] {\scopftab};
  \scLab{1}{0.604}{-4.6pt}{0.60}\scLab{1}{0.390}{4.6pt}{0.39}
  \scLab{2}{3.503}{-4.6pt}{3.50}\scLab{2}{3.757}{4.6pt}{3.76}
  \scLab{3}{6e-7}{0pt}{$6{\times}10^{-7}$}
\end{groupplot}
\node[anchor=south, font=\footnotesize, text=scAxis, inner sep=0pt]
  at ([yshift=6.5mm]$(group c1r1.north west)!0.5!(group c2r1.north east)$) {%
  \tikz[baseline=0pt]\draw[black, line width=0.45pt, fill=scRaw] (0,0) rectangle (2.4mm,2.1mm);\,as is\qquad
  \tikz[baseline=0pt]\draw[black, line width=0.45pt, fill=scFix, postaction={pattern=north east lines, pattern color=scHatch}] (0,0) rectangle (2.4mm,2.1mm);\,after repair\qquad
  \tikz[baseline=0pt]\draw[black, line width=0.45pt, fill=scLoop] (0,0) rectangle (2.4mm,2.1mm);\,Loop-OPF};
\end{tikzpicture}%
%
    \caption{Preventive SC-OPF on the IEEE 300-bus test samples: (a) share of samples meeting the balance, generator, base-flow, and all post-outage flow limits within $10^{-3}$~MW, and (b) cost gap, log scale. Repair: $\ell_1$ repair of the model output.}
    \label{fig:scopf}
\end{figure}
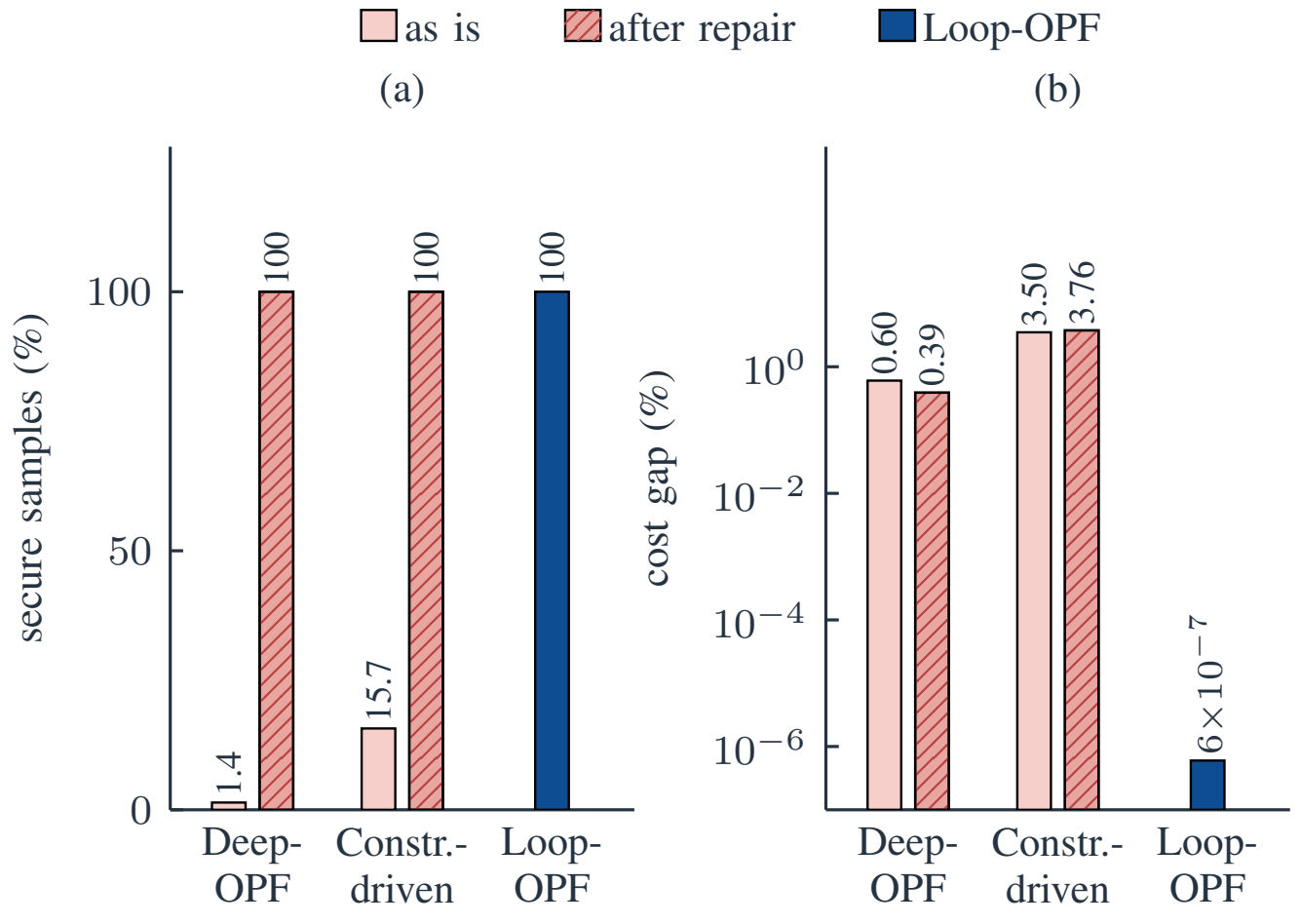

We next test Loop-OPF on the preventive SC-OPF problem{\mohamad~\cite{capitanescu2011scopf}} in Remark~\ref{rem:general}. Here, one dispatch must satisfy the network without outages and every contingency in $\mathcal{C}$ simultaneously. We use the IEEE 300-bus network with a fixed set $\mathcal{C}$ of line outages. As is common in learning for SC-OPF~\cite{pan2019deepopf, park2023pdlscopf}, we exclude the outages that island part of the network. Among the rest, we keep the $32$ lines with the largest base-case flows{\mohamad, a simple form of contingency selection~\cite{ejebe1979contingency}}. For this study, we generate a separate dataset of $10{,}000$ samples. It is again split into $7{,}000$ for training, $1{,}000$ for validation, and $2{,}000$ for testing. We compare with a DeepOPF model~\cite{pan2019deepopf} trained on solver labels and a constraint-driven model~\cite{giraud2024constraint} trained without them. Both are adapted to this setting and trained for $300$ epochs. Each is tested as is and after an $\ell_1$-repair that restores feasibility.

Fig.~\ref{fig:scopf} shows that Loop-OPF is the only method that is secure on every test sample without repair. The two baselines are secure on only $1.4\%$ and $15.7\%$ of the samples. Their mean worst post-outage flow violations are $18.4$ and $12.7$~MW. Repair makes both baselines secure. However, their cost gaps remain at least $0.39\%$, and their RMSEs at least $5.8$~MW.
Loop-OPF reaches an RMSE of $3.9\times10^{-4}$~MW and a cost gap of
$6\times10^{-7}\%$.

\subsection{Comparison with the Untuned Solver}
\label{sec:results_solver}

{\mohamad To isolate the effect of training, we compare the trained Loop-OPF model with the same model at its initial step sizes. Both use the same checkpoint. Each applies exactly $K$ passes of the layer $\Tmap$, that is, $K$ PDHG iterations, for $K=1,2,4,\dots,8{,}192$. Both are evaluated on $2{,}000$ held-out test samples without outages per network. Fig.~\ref{fig:steppair} shows $K\geq 64$, where the two differ most. The shaded area between the curves is the effect of training. For smaller $K$, both models are still far from the solution. Their errors are nonmonotonic in $K$, and neither is better at every $K$.

On IEEE 300, PEGASE 1354, and PEGASE 9241, the trained curve drops earlier than the initial one. It reaches its final RMSE level with about half the iterations. This happens at $K=2{,}048$ instead of $4{,}096$ on IEEE 300 and PEGASE 9241, and at $K=4{,}096$ instead of $8{,}192$ on PEGASE 1354. The gap between the two curves is largest before the trained curve reaches this level. At $K=512$ on IEEE 300, training lowers RMSE by $20\times$ and overload and cost gap by about $40\times$. At $K=1{,}024$ on PEGASE 9241, it lowers RMSE by $60\times$, overload by $72\times$, and cost gap by $119\times$. At $K=4{,}096$ on PEGASE 1354, the trained model has an RMSE of $3\times10^{-4}$~MW, while the initial one is still at $0.43$~MW. The gain is not uniform. At $K=64$, the trained model is worse on PEGASE 1354 in all three metrics and has a higher cost gap on IEEE 300. At $K=128$, it is worse on PEGASE 9241. On the uncongested PEGASE 2869, the two curves nearly coincide, so training brings no visible gain there.

At the largest budget, $K=8{,}192$, which model is better depends on the network and the metric. The trained model is lower in all three metrics on PEGASE 1354 and 9241, with a $3.2\times$ and $3.8\times$ lower overload, respectively. On IEEE 300, it has a $1.6\times$ lower overload but a slightly higher RMSE. On PEGASE 2869, the two are nearly equal. Overall, learning the step sizes mainly shortens the path to the final error level. It roughly halves the iterations needed on the three congested networks, rather than lowering the error at every $K$.}

\begin{figure}[t]
    \centering
    \def\steppaircsv{figures/fig_steppair/steppair.csv}%
    \definecolor{spLoop}{HTML}{0F4D92}%
\definecolor{spAlt}{HTML}{B64342}%
\definecolor{spAxis}{HTML}{253341}%
\def\steppairW{15.4mm}%
\def\steppairH{16mm}%
\definecolor{spGap}{HTML}{F6CFCB}%
\def\spKmin{64}%
\def\spZeroY{2.5e-9}%
\def\spBreakY{2.5e-8}%
\ifdefined\steppaircsv\else\def\steppaircsv{figures/fig_steppair/steppair.csv}\fi
\pgfplotstableread[col sep=comma]{\steppaircsv}\steppairtab%
\def\spZeroCell{0.0}%
\def\spMakeFlow#1{\pgfplotstablecreatecol[create col/assign/.code={%
    \getthisrow{flow_#1}\spCell
    \ifx\spCell\spZeroCell\pgfkeyslet{/pgfplots/table/create col/next content}\spZeroY
    \else\pgfkeyslet{/pgfplots/table/create col/next content}\spCell\fi}]{plot_flow_#1}\steppairtab}%
\spMakeFlow{learned}\spMakeFlow{initial}%
\def\spGlyph#1#2#3{\tikz[baseline=-0.55ex]{\draw[line width=#3, #1] (0,0) -- (6mm,0);
    \path plot[mark=*, mark options={solid, #2}] coordinates {(3mm,0)};}}%
\begin{tikzpicture}[font=\footnotesize, text=spAxis]
\pgfplotsset{
  sp only/.style={x filter/.append code={%
      \getthisrow{system}{\spRow}\edef\spWant{#1}%
      \getthisrow{K}{\spK}%
      \ifx\spRow\spWant\ifnum\spK<\spKmin\relax\def\pgfmathresult{nan}\fi
      \else\def\pgfmathresult{nan}\fi}},
  sp lrn/.style={sp only=#1, color=spLoop, mark=*, mark size=1.3pt, line width=0.9pt,
      mark options={solid, fill=white, line width=0.55pt}},
  sp ini/.style={sp only=#1, color=spAlt, mark=*, mark size=0.75pt, line width=0.9pt,
      dash pattern=on 0.9pt off 1.3pt, line cap=round, mark options={solid, fill=spAlt}},
  sp rmse/.style={ymin=8e-5, ymax=200, ytick={1e-3,1e-1,10}},
  sp flow/.style={ymin=1e-9, ymax=1e4, ytick={1e-6,1e-2,1e2},
      extra y ticks={\spZeroY}, extra y tick labels={}},
  sp gap/.style={ymin=4e-6, ymax=100, ytick={1e-5,1e-3,1e-1,10}},
}
\def\spBreak{%
  \coordinate (spbk) at ({rel axis cs:0,0} |- {axis cs:1,\spBreakY});
  \fill[white] ([xshift=-1.2pt, yshift=-1.2pt]spbk) rectangle ([xshift=1.2pt, yshift=1.2pt]spbk);
  \draw[spAxis, line width=0.5pt] ([xshift=-2.2pt, yshift=-2.2pt]spbk) -- ([xshift=2.2pt, yshift=-0.4pt]spbk)
                                  ([xshift=-2.2pt, yshift=0.4pt]spbk) -- ([xshift=2.2pt, yshift=2.2pt]spbk);}%
\def\spPair#1#2{%
  \addplot[sp lrn=#1, name path=spL] table[x=K, y=#2_learned] {\steppairtab};
  \addplot[sp ini=#1, name path=spI] table[x=K, y=#2_initial] {\steppairtab};
  \addplot[spGap, draw=none, on layer=axis background] fill between[of=spL and spI];}%
\def\spPairFlow#1{%
  \spBreak
  \addplot[sp lrn=#1, name path=spL] table[x=K, y=plot_flow_learned] {\steppairtab};
  \addplot[sp ini=#1, name path=spI] table[x=K, y=plot_flow_initial] {\steppairtab};
  \addplot[spGap, draw=none, on layer=axis background] fill between[of=spL and spI];}%
\begin{groupplot}[
    group style={group size=4 by 3, horizontal sep=2.2mm, vertical sep=2.6mm,
                 xticklabels at=edge bottom, yticklabels at=edge left},
    scale only axis, width=\steppairW, height=\steppairH,
    set layers,
    xmode=log, log basis x=2, xmin=52, xmax=10000,
    xtick={64,512,4096}, xticklabels={64,512,4k},
    ymode=log,
    tick label style={font=\footnotesize, text=spAxis},
    label style={font=\footnotesize, text=spAxis},
    ylabel style={align=center},
    axis lines*=left,
    axis line style={spAxis, line width=0.6pt},
    tick style={spAxis, line width=0.6pt},
    major tick length=2.5pt, minor tick length=0pt,
    title style={font=\footnotesize, text=spAxis, inner sep=0pt, yshift=0.6mm, align=center},
    unbounded coords=discard, filter discard warning=false,
    clip=false,
]
\nextgroupplot[sp rmse, title={IEEE\\300}, ylabel={(a) RMSE\\(MW)}]
  \spPair{IEEE300}{rmse}
\nextgroupplot[sp rmse, title={PEGASE\\1354}]
  \spPair{PEGASE1354}{rmse}
\nextgroupplot[sp rmse, title={PEGASE\\2869\\[-0.3ex]{\scriptsize(uncongested)}}]
  \spPair{PEGASE2869}{rmse}
\nextgroupplot[sp rmse, title={PEGASE\\9241}]
  \spPair{PEGASE9241}{rmse}
\nextgroupplot[sp flow, ylabel={(b) overload\\(MW)}, extra y tick labels={0}]
  \spPairFlow{IEEE300}
\nextgroupplot[sp flow]
  \spPairFlow{PEGASE1354}
\nextgroupplot[sp flow]
  \spPairFlow{PEGASE2869}
\nextgroupplot[sp flow]
  \spPairFlow{PEGASE9241}
\nextgroupplot[sp gap, ylabel={(c) cost gap\\(\%)}]
  \spPair{IEEE300}{gap}
\nextgroupplot[sp gap]
  \spPair{PEGASE1354}{gap}
\nextgroupplot[sp gap]
  \spPair{PEGASE2869}{gap}
\nextgroupplot[sp gap]
  \spPair{PEGASE9241}{gap}
\end{groupplot}
\node[anchor=north, font=\footnotesize, text=spAxis, inner sep=0pt]
  at ([yshift=-5mm]$(group c1r3.south west)!0.5!(group c4r3.south east)$) {number of PDHG iterations $K$ (log scale)};
\node[anchor=south, font=\footnotesize, text=spAxis, inner sep=0pt]
  at ([yshift=12.6mm]$(group c1r1.north west)!0.5!(group c4r1.north east)$) {%
  \spGlyph{spLoop}{fill=white, draw=spLoop, line width=0.55pt, mark size=1.3pt}{0.9pt}\,trained steps\qquad
  \spGlyph{spAlt, dash pattern=on 0.9pt off 1.3pt, line cap=round}{fill=spAlt, draw=spAlt, mark size=0.75pt}{0.9pt}\,initial steps};
\end{tikzpicture}%
%
    \caption{(a) RMSE, (b) mean worst branch overload, and (c) cost gap of Loop-OPF with trained (solid) and initial (dotted) step sizes, each applying exactly $K$ PDHG iterations, for $K\geq 64$, {\mohamad on \textbf{test samples without outages}}. Shading marks the gap between the two. Exact zeros are drawn below the axis break.}
    \label{fig:steppair}
    \par\vspace{2mm}
    \def\ksweepcsv{figures/fig_ksweep/ksweep.csv}%
    \definecolor{kwLoop}{HTML}{0F4D92}%
\definecolor{kwAxis}{HTML}{253341}%
\def\ksweepW{15.4mm}%
\def\ksweepH{16mm}%
\def\kwKmin{64}%
\def\kwZeroY{2.5e-9}%
\def\kwBreakY{2.5e-8}%
\ifdefined\ksweepcsv\else\def\ksweepcsv{figures/fig_ksweep/ksweep.csv}\fi
\pgfplotstableread[col sep=comma]{\ksweepcsv}\ksweeptab%
\def\kwZ#1{\thisrow{#1}==0 ? \kwZeroY : \thisrow{#1}}%
\def\kwBreak{%
  \coordinate (kwbk) at ({rel axis cs:0,0} |- {axis cs:1,\kwBreakY});
  \fill[white] ([xshift=-1.2pt, yshift=-1.2pt]kwbk) rectangle ([xshift=1.2pt, yshift=1.2pt]kwbk);
  \draw[kwAxis, line width=0.5pt] ([xshift=-2.2pt, yshift=-2.2pt]kwbk) -- ([xshift=2.2pt, yshift=-0.4pt]kwbk)
                                  ([xshift=-2.2pt, yshift=0.4pt]kwbk) -- ([xshift=2.2pt, yshift=2.2pt]kwbk);}%
\begin{tikzpicture}[font=\footnotesize, text=kwAxis]
\pgfplotsset{
  kw kmin/.style={x filter/.append code={%
      \getthisrow{k}{\kwK}\ifnum\kwK<\kwKmin\relax\def\pgfmathresult{nan}\fi}},
  kw line/.style={kw kmin, color=kwLoop, mark=*, mark size=1.3pt, line width=0.9pt,
      mark options={solid, fill=white, line width=0.55pt}},
  kw rmse/.style={ymin=8e-5, ymax=200, ytick={1e-3,1e-1,10}},
  kw flow/.style={ymin=1e-9, ymax=1e4, ytick={1e-6,1e-2,1e2},
      extra y ticks={\kwZeroY}, extra y tick labels={}},
  kw gap/.style={ymin=4e-6, ymax=100, ytick={1e-5,1e-3,1e-1,10}},
}
\begin{groupplot}[
    group style={group size=4 by 3, horizontal sep=2.2mm, vertical sep=2.6mm,
                 xticklabels at=edge bottom, yticklabels at=edge left},
    scale only axis, width=\ksweepW, height=\ksweepH,
    xmode=log, log basis x=2, xmin=52, xmax=10000,
    xtick={64,512,4096}, xticklabels={64,512,4k},
    ymode=log,
    tick label style={font=\footnotesize, text=kwAxis},
    label style={font=\footnotesize, text=kwAxis},
    ylabel style={align=center},
    axis lines*=left,
    axis line style={kwAxis, line width=0.6pt},
    tick style={kwAxis, line width=0.6pt},
    major tick length=2.5pt, minor tick length=0pt,
    title style={font=\footnotesize, text=kwAxis, inner sep=0pt, yshift=0.6mm, align=center},
    unbounded coords=discard, filter discard warning=false,
    clip=false,
]
\nextgroupplot[kw rmse, title={IEEE\\300}, ylabel={(a) RMSE\\(MW)}]
  \addplot[kw line] table[x=k, y=rmse300] {\ksweeptab};
\nextgroupplot[kw rmse, title={PEGASE\\1354}]
  \addplot[kw line] table[x=k, y=rmse1354] {\ksweeptab};
\nextgroupplot[kw rmse, title={PEGASE\\2869\\[-0.3ex]{\scriptsize(uncongested)}}]
  \addplot[kw line] table[x=k, y=rmse2869] {\ksweeptab};
\nextgroupplot[kw rmse, title={PEGASE\\9241}]
  \addplot[kw line] table[x=k, y=rmse9241] {\ksweeptab};
\nextgroupplot[kw flow, ylabel={(b) overload\\(MW)}, extra y tick labels={0}]
  \kwBreak
  \addplot[kw line] table[x=k, y expr=\kwZ{flow300}] {\ksweeptab};
\nextgroupplot[kw flow]
  \kwBreak
  \addplot[kw line] table[x=k, y expr=\kwZ{flow1354}] {\ksweeptab};
\nextgroupplot[kw flow]
  \kwBreak
  \addplot[kw line] table[x=k, y expr=\kwZ{flow2869}] {\ksweeptab};
\nextgroupplot[kw flow]
  \kwBreak
  \addplot[kw line] table[x=k, y expr=\kwZ{flow9241}] {\ksweeptab};
\nextgroupplot[kw gap, ylabel={(c) cost gap\\(\%)}]
  \addplot[kw line] table[x=k, y=gap300] {\ksweeptab};
\nextgroupplot[kw gap]
  \addplot[kw line] table[x=k, y=gap1354] {\ksweeptab};
\nextgroupplot[kw gap]
  \addplot[kw line] table[x=k, y=gap2869] {\ksweeptab};
\nextgroupplot[kw gap]
  \addplot[kw line] table[x=k, y=gap9241] {\ksweeptab};
\end{groupplot}
\node[anchor=north, font=\footnotesize, text=kwAxis, inner sep=0pt]
  at ([yshift=-5mm]$(group c1r3.south west)!0.5!(group c4r3.south east)$) {number of PDHG iterations $K$ (log scale)};
\end{tikzpicture}%
%
    \caption{(a) RMSE, (b) mean worst branch overload, and (c) cost gap of Loop-OPF against $K$.  {\mohamad Same layout and axes as Fig.~\ref{fig:steppair}, for $K\geq 64$, on the $2{,}000$ \textbf{test samples with and without outages} per network, with the models trained on both. Exact zeros are drawn below the axis break.}}
    \label{fig:budget}
\end{figure}

\subsection{Inference Budget}
\label{sec:results_fixedk}

A key advantage of Loop-OPF is that it is an \textit{anytime}
method~\cite{zilberstein1996anytime}. It can return a dispatch at any point during inference. The dispatch then improves smoothly as more computation is spent. At inference, as many layers $\Tmap$ can be applied as the time budget
permits, with no retraining. In contrast, the feed-forward predictors of
Section~\ref{sec:results_clean} have a fixed depth. We test this flexibility
by varying the number $K$ of layers applied at inference.

Fig.~\ref{fig:budget} shows a progressive improvement in dispatch accuracy,
flow feasibility, and cost as the iteration budget increases. The model can
return a dispatch at a short budget and refine it when more computation is
available. {\mohamad At $K=8{,}192$, the RMSE is below $0.1$~MW and the mean worst flow violation is below $0.21$~MW on every network. On PEGASE 9241, they are $0.014$ and $0.031$~MW, with a cost gap of $6\times10^{-4}\%$.} %
 On {\mohamad IEEE 300, PEGASE 1354, and PEGASE 2869}, each iteration takes about $1.55$ to $1.95$~ms.

\noindent{\mohamad \textbf{Summary.} The tests lead to four main findings. First, on samples without outages, Loop-OPF has the lowest RMSE on all four networks, at most $0.051\%$ of the average generator output. Second, under line and generator outages unseen in training, it outperforms models that receive the post-outage topology as input. Its flow violation is at most $1.6\%$, against up to $17.6\%$ for these models. Third, in the preventive SC-OPF, it is the only method that meets every post-outage flow limit without repair. Fourth, on the congested networks, the trained Loop-OPF needs roughly half the iterations of the untrained one, and it can return a usable dispatch at any iteration budget. Each iteration takes only about $1.55$ to $1.95$~ms on IEEE 300, PEGASE 1354, and PEGASE 2869.}
\section{Conclusion}
\label{sec:conclusion}

This paper develops Loop-OPF, a solver-informed neural network for solving the OPF problem with guaranteed optimality at convergence and inference-time topology adaptivity. Each layer of Loop-OPF performs one PDHG iteration, so that its fixed point is equivalent to the optimality condition and thus attains the global optimum of convex OPF problems. To accelerate its convergence, Loop-OPF learns the PDHG step-size parameters, which are reparameterized to remain admissible throughout training. Furthermore, the DEQ framework with implicit differentiation enables self-supervised training without OPF solution labels, with a training memory that does not grow with the number of layers. Going beyond a fixed topology, Loop-OPF incorporates the post-contingency network  into every layer through a simple update, such that the same trained Loop-OPF applies to unseen line and generator outages without retraining. Numerical tests on the IEEE 300-bus and the PEGASE 1354-, 2869-, and 9241-bus networks have validated that the learned parameters improve the accuracy and flow feasibility of Loop-OPF, outperforming state-of-the-art learning-for-OPF methods with the smallest cost gap and line-flow violations. More importantly, Loop-OPF maintains the accuracy and feasibility for any new OPF scenario under line or generator outages, achieving the inference-time adaptivity.

{\mohamad Exciting future directions open up for this work. First, one can learn scenario-specific step-size parameters, so that each new OPF scenario predicts its own step sizes through a meta-learning framework. Second, since PDHG relies only on matrix--vector products, Loop-OPF could be implemented in a distributed manner across areas or control centers. Last, the feedback structure of Loop-OPF makes it attractive for optimal control designs where the control rules affect the network equilibrium, with notable applications in power systems, transportation networks and other cyber-physical systems.}

\bibliographystyle{IEEEtran}
\itemsep2pt
\bibliography{refs}

\end{document}